\documentclass[lettersize,journal]{IEEEtran}
\usepackage{amsmath,amsfonts}
\usepackage{algorithmic}
\usepackage{algorithm}
\usepackage{array}
\usepackage[caption=false,font=normalsize,labelfont=sf,textfont=sf]{subfig}
\usepackage{textcomp}
\usepackage{stfloats}
\usepackage{url}
\usepackage{verbatim}
\usepackage{graphicx}
\usepackage{cite}
\usepackage[most]{tcolorbox}
\tcbset{width=0.48\textwidth,boxrule=1pt,colback=lightgray,arc=1pt,auto outer arc,left=0pt,right=0pt,top=0pt,bottom=0pt,boxsep=2pt}
\usepackage{subcaption}
\usepackage{multirow}
\usepackage{amsmath}
\usepackage{amssymb}
\usepackage{url}
\usepackage{booktabs}
\usepackage{bbding}
\tcbuselibrary{breakable}
\usepackage{makecell}
\usepackage{cleveref}
\usepackage{colortbl}
\definecolor{gyellow}{HTML}{F4B400}
\definecolor{mildyellow}{HTML}{FFF2CC}
\definecolor{whitesmoke}{HTML}{EEE9E9}
\definecolor{deepred}{HTML}{000000}

\newcommand{\mypara}[1]{\medskip\noindent{\bf {#1}.}}

\newcommand{\xl}[1]{\boldsymbol{#1}}
\usepackage{color}

\usepackage{xspace} 
\usepackage{bbding}
\newcommand{\leak}{\textit{LoRA-Leak}\xspace}

\begin{document}

\title{\leak: Membership Inference Attacks Against LoRA Fine-tuned Language Models}

\author{
Delong Ran$^{1}$, Xinlei He$^{2}$, Tianshuo Cong$^{1}$\textsuperscript{(}\Envelope\textsuperscript{)}, Anyu Wang$^{1}$, Qi Li$^{1}$, and Xiaoyun Wang$^{1}$
\\
\textsuperscript{\rm 1}Tsinghua University,
\textsuperscript{\rm 2}Hong Kong University of Science and Technology (Guangzhou)
\thanks{
\textsuperscript{(}\Envelope\textsuperscript{)}Correspondence to: Tianshuo Cong (congtianshuo@gmail.com).
}
}

\markboth{Journal of \LaTeX\ Class Files,~Vol.~14, No.~8, August~2021}%
{Shell \MakeLowercase{\textit{et al.}}: A Sample Article Using IEEEtran.cls for IEEE Journals}

\maketitle

\begin{abstract}
Language Models (LMs) typically adhere to a ``pre-training and fine-tuning'' paradigm, where a universal pre-trained model can be fine-tuned to cater to various specialized domains. Low-Rank Adaptation (LoRA) has gained the most widespread use in LM fine-tuning due to its lightweight computational cost and remarkable performance. 
Because the proportion of parameters tuned by LoRA is relatively small, there might be a misleading impression that the LoRA fine-tuning data is invulnerable to Membership Inference Attacks (MIAs).
However, we identify that utilizing the pre-trained model can induce more information leakage, which is neglected by existing MIAs.
Therefore, we introduce \leak, a holistic evaluation framework for MIAs against the fine-tuning datasets of LMs.
\leak incorporates fifteen membership inference attacks, including ten existing MIAs, and five improved MIAs that leverage the pre-trained model as a reference.
In experiments, we apply \leak to three advanced LMs across three popular natural language processing tasks, demonstrating that LoRA-based fine-tuned LMs are still vulnerable to MIAs (e.g., 0.775 AUC under conservative fine-tuning settings).
We also applied \leak to different fine-tuning settings to understand the resulting privacy risks.
We further explore four defenses and find that only dropout and excluding specific LM layers during fine-tuning effectively mitigate MIA risks while maintaining utility.
We highlight that under the ``pre-training and fine-tuning'' paradigm, the existence of the pre-trained model makes MIA a more severe risk for LoRA-based LMs. 
We hope that our findings can provide guidance on data privacy protection for specialized LM providers.
\end{abstract}


\section{Introduction}
\IEEEPARstart{L}{anguage} Models (LMs) have been extensively utilized in a variety of Natural Language Processing (NLP) tasks, including legal advise~\cite{cheng2024adapting}, scientific research~\cite{nguyen-etal-2023-astrollama}, etc. 
Despite the belief that pre-trained LMs such as ChatGPT~\cite{chatgpt} and Llama~\cite{touvron2023llama} have exhibited the rudiments of Artificial General Intelligence (AGI), their data-driven nature results in sub-optimal performance in specialized domains~\cite{roziere2023code}. 
To address this issue, the general paradigm for tailoring LMs to downstream tasks consists of two steps: \textit{pre-training} and \textit{fine-tuning}.
The pre-training process aims to learn rich language features and structures from a massive corpus in an unsupervised way, forming the \textit{pre-trained models} capable of mastering general language patterns.
Consequently, these pre-trained models can serve as a remarkable starting point and further fine-tuned on vertical domains in a supervised way, resulting in the \textit{specialized models} that are adept at diverse downstream tasks.
For example, codeLlama~\cite{roziere2023code} and AstroLLaMA~\cite{nguyen2023astrollamaspecializedfoundationmodels} are the fine-tuned variants of Llama-2~\cite{touvron2023llama} on programming domain and astronomy domain, respectively. 

\begin{figure}[t]
\centering
\includegraphics[width=0.9\linewidth]{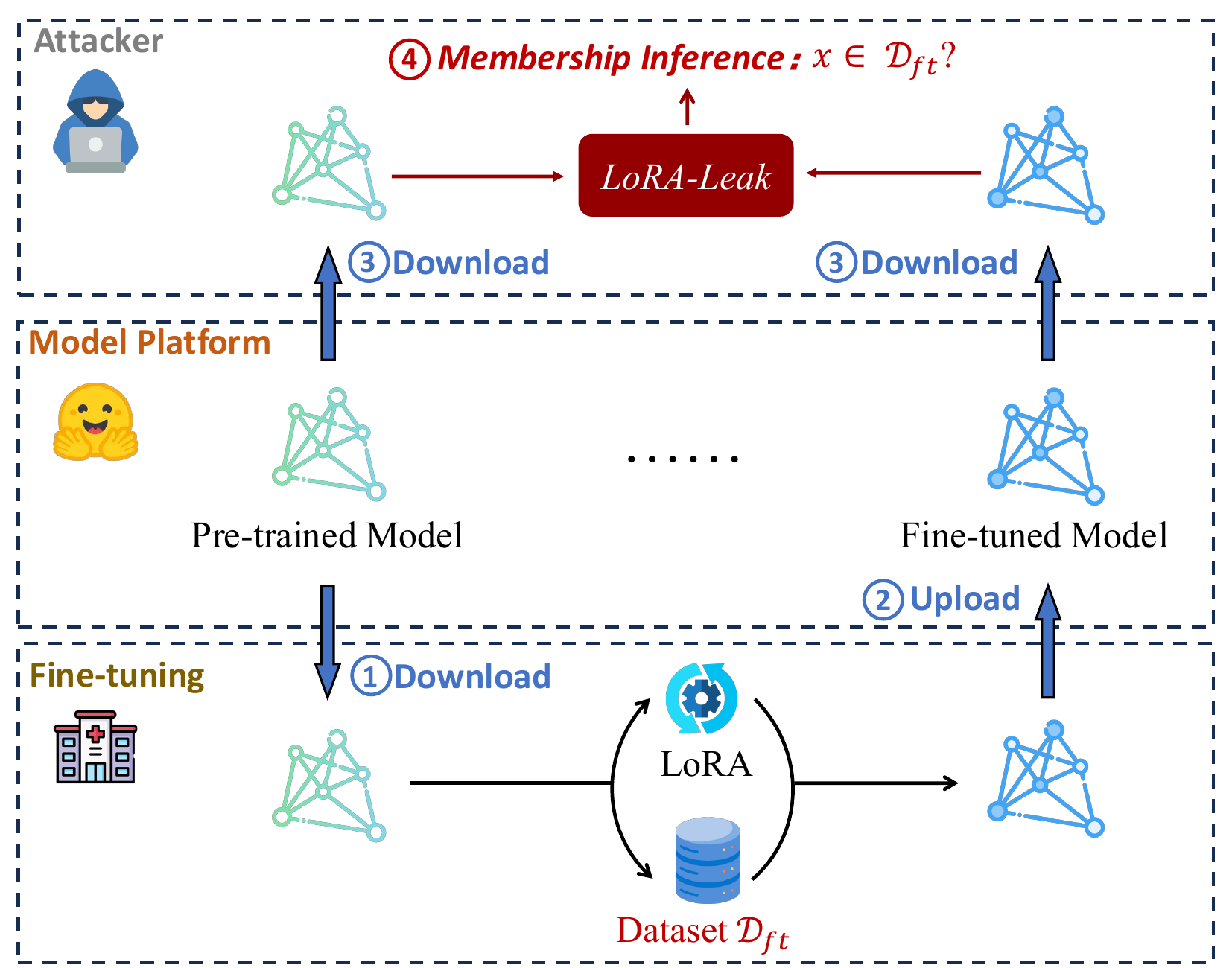}
\caption{Overview of \leak. \leak aggregates information from the specialized fine-tuned model and its pre-trained model to launch more powerful MIAs against LMs.}
\label{fig:overview}
\end{figure}

The performance of the fine-tuned specialized LMs hinges on two key factors: the fine-tuning algorithms and the fine-tuning datasets (see~\Cref{fig:overview}).
Given the current scale of parameters in LMs, the computational cost of full-parameter fine-tuning is prohibitively expensive (e.g., a 16-bit full-tuning for Llama-7B requires 60GB of GPU memory~\cite{zheng2024llamafactory}). 
This limitation has spurred the rapid development of Parameter-Efficient Fine-Tuning (PEFT)~\cite{peft}.
Specifically, Low-Rank Adaptation (LoRA)~\cite{hu2022lora} is the state-of-the-art (SOTA) Parameter-Efficient Fine-Tuning (PEFT) framework with the highest usage. By the end of 2023, there were over 12,000 LoRA models on Hugging Face, with some receiving over one  hundred thousand downloads per month~\cite{dong2024philosophersstonetrojaningplugins}.
Since LoRA only trains side-loaded rank-decomposition matrices while keeping the model backbones frozen, it only needs 16GB to fine-tune Llama-7B, which can be further reduced to 6GB through its 4-bit quantization version qLoRA~\cite{dettmers2023qlora}.
Moreover, a high-quality fine-tuning dataset is also paramount as it serves as the specialized knowledge source and determines the upper limit of the model's capability.
Considering the potential presence of privacy-sensitive information within fine-tuning datasets, such as those in financial and medical domains, a comprehensive assessment for privacy leakage associated with fine-tuning datasets is of vital importance.

Membership Inference Attacks (MIAs)~\cite{9833649}, in which attackers aim to determine if a specific sample was part of the training data of a target model, pose a persistent privacy threat to machine learning models. 
Recently, with the emergence of an increasing number of specialized LMs in the open-source model zoo (e.g., Huggingface\footnote{\url{https://huggingface.co/models}.}), conducting MIAs against the fine-tuning datasets of the LoRA-based fine-tuned LMs has become a prominent research focus.
Since LoRA only fine-tunes a small subset of the model's parameters, recent studies suggest that their fine-tuning datasets are invulnerable to MIA ~\cite{wen2023standing,liu2024precurious}.
However, the current research fails to recognize that the publicly accessible pre-trained model could introduce additional privacy threats. 

\mypara{Our Work}
We propose \leak, a holistic evaluation framework for measuring the vulnerability of LoRA-based fine-tuned LMs against MIAs.
To comprehensively assess the privacy risks of the LoRA fine-tuning dataset, we formulate three \textit{Research Questions} (\textit{\textbf{RQs}}):

\begin{itemize}
\item \textit{\textbf{RQ1}}: Is MIA still a serious privacy threat for LoRA-based fine-tuned LMs?
\item \textit{\textbf{RQ2}}: Can incorporating pre-trained model information lead to the design of more potent MIAs?
\item \textit{\textbf{RQ3}}: What LoRA fine-tuning strategies can mitigate the threat of MIAs?
\end{itemize}

To answer \textit{\textbf{RQ1}}, we propose \leak, a comprehensive framework for MIA against LoRA finetuning, incorporating fifteen diverse attack methods (see~\Cref{tab:meth}).
Subsequently, we assess the effectiveness of MIAs in \leak against nine LoRA fine-tuned LMs, developed using three widely used LMs and three practical fine-tuning datasets under a conservative setting to prevent overfitting.
Our experimental results demonstrate that \leak can achieve high AUC scores against LoRA-based fine-tuned LMs.
For instance, the AUC scores against Llama-2 model fine-tuned on AG News~\cite{NIPS2015_250cf8b5}, OAsst~\cite{10.5555/3666122.3668186}, and MedQA~\cite{jin2020disease} are 0.765, 0.721, and 0.775, respectively.

To demonstrate the necessity of introducing pre-trained models for answering \textit{\textbf{RQ2}}, we compare the effectiveness of different MIAs the performance of various MIAs with and without using the pre-trained model as a reference.
We observe that the calibration from the pre-trained model can consistently amplify the privacy risk (see~\Cref{tab:auc_llama}).
For a more in-depth analysis, we discuss the impact of different kinds of reference models~\cite{9833649}.
As~\Cref{fig:ref-model-discuss} shows, introducing other kinds of reference models cannot achieve the optimal attack results as introducing pre-trained models.

To address \textit{\textbf{RQ3}}, we comprehensively discuss various fine-tuning settings.
We first analyze the influence of fine-tuning hyperparameters on the attack effect, such as the fine-tuning epoch and the selection of LoRA fine-tuning modules.
Furthermore, we discuss four potential defenses.
We first explore three traditional defense strategies, i.e., dropout, weight decay, and differential privacy (DP), in which only dropout can mitigate the risk of MIAs while preserving utility.
Furthermore, in~\Cref{sec:def-exc}, we demonstrate that fine-tuning excluding specific modules can also mitigate privacy risks.

In summary, our contributions are as follows:
\begin{itemize}
    \item We introduce \leak, a comprehensive evaluation framework on MIAs against LoRA-based fine-tuned LMs.
    \item We propose that introducing pre-trained models into the inference attacking pipeline can effectively amplify the privacy risks.
    \item We explore four defenses and find that dropout and fine-tuning excluding specific layers can mitigate the threat of \leak.
\end{itemize}

\section{Preliminaries}

\subsection{Causal Language Model (CLM)}

A causal language model (CLM) is designed for next-token prediction tasks over a token space $\mathcal{T}$.
It first takes a sequence of tokens $x_{1:n} \in \mathcal{T}^n$ as input and transforms each token $x_i$ into a continuous embedding $\boldsymbol{e}_i$. These embeddings are then fed into a decoder-only transformer $\mathcal{M}$, which produces the probability of each token $x_{n+1} \in \mathcal{T}$ being the next token, i.e., $\boldsymbol{p}_n = \langle \Pr(x_{n+1}|x_{1:n})\rangle_{x_{n+1}\in \mathcal T}$.

The primary purpose of such a model is to generate reasonable text completions for user inputs by repeatedly performing next-token predictions.
This is achieved by training the model on a collection of token sequences in an autoregressive manner:
For each training sample $x_{1:n} \in \mathcal{T}^n$, the objective is to minimize the model's perplexity (PPL) on that sequence, defined in the form of average cross-entropy loss as
\begin{equation}
\label{ppl}
\mathcal{L}(x_{1:n};\mathcal M)=-\frac{1}{n-1}\sum_{i=1}^{n-1}\log{p_{i,x_{i+1}}}.
 \end{equation}

\subsection{Low-Rank Adaptation (LoRA)}

Low-Rank Adaptation (LoRA) is one of the most widely used Parameter-Efficient Fine-Tuning (PEFT) algorithms for model fine-tuning~\cite{dong2024philosophersstonetrojaningplugins}.
For a pre-trained model $\mathcal M_{pt}$, this method selects only a subset of layers for fine-tuning. 
For each selected layer, it freezes the pre-trained weight $W_i \in \mathbb{R}^{d \times k}$ and introduces two additional decomposition matrices $\Delta_i = (A_i, B_i) \in \mathbb{R}^{d \times r} \times \mathbb{R}^{r \times k}$ to fine-tune, where $r \ll \min (d,k)$ is the hyperparameter of rank.
In the resulting fine-tuned model $\mathcal M_{ft}$, its layer is then represented as:
\begin{equation}
\label{eq:lora}
W'_{i} = W_{i} + A_i B_i.
\end{equation}

\subsection{Membership Inference Attack}
\label{sec:mia-def}
Membership inference (MI) is a privacy game where an adversary $\mathcal A$, given access to a machine learning model $\mathcal M$, aims to determine whether a specific record $x$ is part of the model's training dataset $\mathcal D$, i.e., $\mathcal A(x;\mathcal M)\rightarrow \{0,1\}$.
The adversary wins if and only if $\mathcal A(x;\mathcal M)=\mathbb{I}[x\in\mathcal D]$.
To achieve this, the adversary will choose a score function $\mathcal S(x;\mathcal M)\rightarrow \mathbb R$ and a threshold $\tau \in \mathbb R$.
Finally, the adversary determines membership based on the rule $\mathcal A(x;\mathcal M)=\mathbb{I}[S(x;\mathcal M)>\tau]$.

\begin{figure*}[t]
\centering
\includegraphics[width=0.88\linewidth]{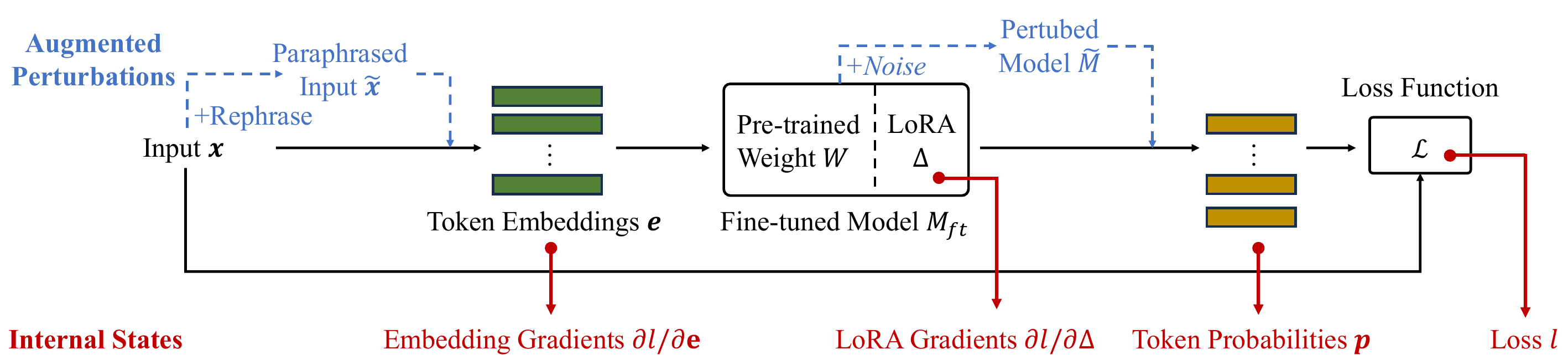}
\caption{
The pipeline of \leak.
The sample $x$ is fed to the target model to infer its membership.
During the forward and back propagation, the internal states of the model can be perturbed as the dotted line indicates.
Some internal states of the model can be extracted to initiate attacks as the red arrow indicates.
These signals can be further calibrated by pre-trained models to obtain more effective MIAs.
}
\label{fig:pipeline}
\end{figure*}

\begin{table*}[t]
\caption{The list of fifteen MIAs integrated within \leak. \leak incorporates eight well-established MIAs, and refines six of these leveraging the pre-trained model as a reference.}
\label{tab:meth}
\centering
\begin{tabular}{l|c|c|c|c}
\toprule
\textbf{Attack} & \textbf{Internal State} & \textbf{Augmented Perturbation} & \textbf{Used to attack LoRA's $\mathcal D_{ft}$?} & \textbf{Calibrated by referencing $\mathcal{M}_{pt}$?}\\ \midrule \midrule
LOSS \cite{jagannatha2021membershipinferenceattacksusceptibility} & $l$ & - & $\checkmark$ (\hspace{1sp}\cite{wen2023standing}) & $\checkmark$ (\hspace{1sp}\cite{mireshghallah-etal-2022-quantifying}) \\ 
zlib~\cite{carlini2021extracting} & $l$ & - & $\times$ & $\times$ (already calibrated by zlib)\\  
Neighborhood \cite{mattern-etal-2023-membership} & $l$ & $ \tilde x$ & $\checkmark$ (\hspace{1sp}\cite{fu2023practical})&$\checkmark$ \textbf{(Ours)}\\ 
SPV~\cite{fu2023practical} & $l$ & - & \checkmark & $\times$ (already calibrated by $\mathcal{M}_{sp}$) \\
MoPe~\cite{li-etal-2023-mope} & $l$ & $\tilde{ \mathcal{M}_{ft}}$ & $\times$ &$\checkmark$ \textbf{(Ours)}\\ 
Min-K\%~\cite{shi2023detecting} & $h$ & - & $\times$ &$\checkmark$ \textbf{(Ours)}\\ 
Min-K\%++~\cite{zhang2024min} & $h$ & - & $\times$ &$\checkmark$ \textbf{(Ours)}\\ 
$\text{GradNorm}_\theta$~\cite{wang2024pandoras} & ${\partial l}/{\partial \Delta}$ & - & $\times$ &$\times$ (no correspondence for $\mathcal{M}_{pt}$) \\ 
$\text{GradNorm}_x$~\cite{wang2024pandoras} & ${\partial l}/{\partial e}$ & - & $\times$ &$\checkmark$ \textbf{(Ours)}\\ 
\bottomrule
\end{tabular}
\end{table*}

\section{Related Work}

\subsection{MIAs against Neural Networks}
Membership inference attack (MIA) was initially proposed to detect the training sample of image classification tasks in artificial neural networks~\cite{SSSS17}.
It utilizes shadow datasets to train multiple shadow models as proxies to mimic the behavior of the target model.
The prediction results of these shadow models are then collected to train a binary classifier, which learns the characteristics of membership samples.
Nasr et al.~\cite{NSH19} extend this approach to the white-box setting, where the model's internal gradients are also available.
Based on this additional feature, the classifier could achieve a more accurate result.
However, as the model's complexity increases, it becomes impractical to select effective and learnable features for training the classifier.

Another approach is to identify an explicit intermediate signal from the model to determine membership.
For example, using the prediction correctness~\cite{255348} or prediction confidence~\cite{10.1145/3319535.3354211} as a metric for membership.
The Likelihood Ratio Attack (LiRA)~\cite{9833649} is one of the most effective metric-based MIAs, which calibrates the membership signal by comparing the prediction confidence between the target model and the shadow models.
We extend the idea to develop MIAs against fine-tuned language models.

Despite proposing various attacks, there is also theoretical research on the mechanics of membership inference.
Yeom et al.~\cite{8429311} highlight that the primary threat of membership inference is due to overfitting.
Additionally, Bentley et al.~\cite{bentley2020quantifyingmembershipinferencevulnerability} further quantitatively associate the risk of membership inference with the target model's generalization gap.
Therefore, we report the model's generalization gap to measure its inherent vulnerability to membership inference.

\subsection{MIAs against Pre-trained Language Models}
With the advent of Pre-trained Language Models (PLM), the membership inference attacks against their pre-training corpora are surpassing.
Earlier research primarily adapted from the MIAs against neural networks, such as the LOSS attack~\cite{jagannatha2021membershipinferenceattacksusceptibility} and the LiRA attack~\cite{mireshghallah-etal-2022-quantifying}.
Recently, several attacks that specifically exploit the characteristics of LMs have been proposed.
For example, the difference of token probabilities inspires Min-K\%~\cite{shi2023detecting} and Min-K\%++~\cite{zhang2024min}, while MoPe~\cite{li-etal-2023-mope} exploits the smoothness around the training points.
However, the credibility of these attacks is hindered by their faulty evaluation method~\cite{meeus2024sokmembershipinferenceattacks}.
Because PLMs lack transparency regarding their training corpus, these works use corpora from before and after the model's knowledge cut-off date as a proxy for ground truth.
Consequently, there is a distribution shift between members and non-members, allowing even classifiers without access to the model to achieve high accuracy~\cite{das2024blindbaselinesbeatmembership}.
As for the benchmark with a fair membership split, these attacks degrade to a nearly random performance against PLMs~\cite{duan2024membershipinferenceattackswork}.
Therefore, we explore whether those attacks against PLMs can be effective against the fine-tuning corpus in LoRA fine-tuning under a fair membership split.

\subsection{MIA against Fine-tuned Language Models}
There are also MIAs against LLM's fine-tuning samples.
The neighborhood attack~\cite{mattern-etal-2023-membership} calibrates the loss signal by the average loss of rephrased samples.
MIA-SPV~\cite{fu2023practical} further enhances this attack with LiRA by comparing the calibrated loss from the target model with the calibrated loss from a self-prompted shadow model.
Although this attack has been reported to achieve good performance, it is important to note that their target models were trained for 10 epochs, making them inherently vulnerable to membership inference (See~\Cref{sec:eval-main}).
Consequently, many MIAs could achieve similar performance, rendering the advantage of this method unclear.
Moreover, this attack is computational-intensive for generating self-prompt samples.

Although existing MIAs against fine-tuned LLMs are delicate and promising, they often overlook the significance of the pre-trained model.
The reference attack~\cite{mireshghallah2022memorizationnlpfinetuningmethods}, a simple yet effective approach, draws inspiration from the LiRA attack by using the pre-trained model as a shadow model to calibrate the loss signal of the target model.
This suggests that referencing pre-trained models could further amplify the privacy risks associated with LoRA fine-tuned LMs.
Therefore, we extend this concept to fill the gaps in all existing MIAs by calibrating their signals with their pre-trained model.

LoRA has already been the most used language model fine-tuning algorithm~\cite{dong2024philosophersstonetrojaningplugins}, but the studies for its privacy risks are still incomplete.
Wen et al.~\cite{wen2023standing} reported that LoRA fine-tuning is invulnerable to MIAs.
However, their work only employs the LiRA attack~\cite{mireshghallah-etal-2022-quantifying}, which may not fully reveal the MIA risks of LoRA fine-tuning.
In our work, we comprehensively explore the privacy risks evoked by LoRA with fifteen MIAs against different fine-tuning settings and defense strategies.

Recently, Liu et al.~\cite{liu2024precurious} proposed \textit{PreCurious}, a framework designed to amplify membership inference risks in fine-tuned language model by poisoning its pre-trained model.
However, their threat model is strong because the victim must use the corrupted model provided by the adversary.
In this work, we assume the victim uses the official open-source pre-trained model that is also accessible to the adversary.
This is a more practical threat model in LoRA fine-tuning scenarios that still significantly amplifies privacy risks.


\section{LoRA-Leak}


In this section, we introduce \leak, a comprehensive framework for MIAs against LoRA fine-tuning.
We begin by outlining the threat model of \leak.
Next, we present a systematic taxonomy that identifies the essential components of MIAs to categorize all existing MIAs within this framework. 
This framework proposes a neglected attacking surface that utilizing the pre-trained model as a reference could further enhance existing MIAs.

\subsection{Threat Model}
\label{sec:threat}

In our threat model, the victim is a model fine-tuner aiming to build a specialized CLM $\mathcal{M}_{ft}$ for a downstream task using their private dataset $\mathcal{D}_{ft}$.
To achieve this, the victim first obtains a renowned PLM $\mathcal{M}_{pt}$ that is publicly released by a benign party, such as OpenAI's GPT-2 or Meta's Llama-2.
Subsequently, the victim fine-tunes $\mathcal{M}_{pt}$ on $\mathcal{D}_{ft}$ using LoRA to get the resulting model $\mathcal{M}_{ft}$.
Finally, the victim publicly releases the fine-tuned LoRA model $\mathcal{M}_{ft}$ in a model zoo such as Hugging Face.

\mypara{Adversary's Goal}
The adversary's goal is to infer whether a record $x$ belongs to the fine-tuning dataset $\mathcal{D}_{ft}$ of the target model $\mathcal{M}_{ft}$.
Note that the adversary aims to infer the membership in the victim's private fine-tuning dataset $\mathcal{D}_{ft}$, rather than the pre-training dataset that is used to train $\mathcal{M}_{pt}$.

\mypara{Adversary's Knowledge} 
The adversary has full knowledge of the final fine-tuned LoRA model $\mathcal{M}_{ft}$ but does not know its fine-tuning details such as the fine-tuning hyperparameters.
The attacker also does not know any information about the fine-tuning or pre-training datasets, even their domains.
However, we additionally assume the adversary has full knowledge of the pre-trained model from which the target model was fine-tuned.
This assumption is based on the fact that LoRA models must be used with their pre-trained model.
Consequently, the name of the pre-trained model is typically specified in the target LoRA model's metadata or model card, allowing the adversary to effortlessly obtain this PLM by referencing the name.

\mypara{Adversary's Capability}
We assume that the adversary possesses sufficient GPU resources for model inference and backpropagation.
Since the adversary fully possesses the fine-tuned model and its pre-trained model, that means they can self-host these models.
As a result, the adversary has white-box access to the models, including all internal states during inference, such as sample loss, predicted token probabilities, and gradients, etc.
However, the adversary cannot interfere with the pre-training and fine-tuning process, nor can they poison the victim's pre-trained model or fine-tuning dataset, as these actions typically require sophisticated supply-chain attacks.
Some literature~\cite{fu2023practical} proposed a gray-box scenario, where the adversary is limited to access partial internal states, such as the loss of the sample or prediction probabilities of each token.
While these attacks appear to operate in a more constrained scenario, current inference APIs, such as OpenAI's Developer Platform\footnote{\url{https://platform.openai.com/docs/api-reference/chat/create}} and Hugging Face's Inference Endpoints\footnote{\url{https://huggingface.co/docs/inference-endpoints/index}}, do not provide any internal states that aligns their assumption.
Therefore, we focus on the white-box scenario to fully expose the threat of membership inference in a passive setting.

\subsection{Holistic Framework for MIAs against LMs}
As discussed in \Cref{sec:mia-def}, the essence of MIA lies in selecting an appropriate score function $\mathcal S(x;\mathcal M)$ that effectively differentiates between members and nonmembers.
Here, we identify the key components involved in the design of $\mathcal S(x;\mathcal M)$, including \textit{Intermediate States}, \textit{Augmented Perturbations}, and \textit{Referenced Calibrations}.
We will demonstrate how this holistic framework can encompass existing MIAs and lead to our proposed enhancements.

\mypara{Internal States}
To calculate $\mathcal S(x;\mathcal M)$, the text sample $x$ is fed into the fine-tuned model $\mathcal{M}$ for forward or backpropagation.
During this process, the adversary can collect various  \textit{internal states} as their knowledge base for initiating the attack.
As illustrated in~\Cref{fig:pipeline}, there are four internal states that correlate with the sample's membership status.
The \textit{loss of sample} is denoted as $\mathcal L(x;M)$.
Empirically, members tend to have smaller loss value than nonmembers, leading to the well-known LOSS attack~\cite{jagannatha2021membershipinferenceattacksusceptibility}, whose score function is defined as 
\begin{equation}
\label{loss}
S_{loss}(x;\mathcal M)=-\mathcal L(x;M).
\end{equation}
The \textit{predicted next-token probabilities for position $i$} are denoted as $\xl p_{i}=\langle \Pr(x_{i+1}|x_{1:i})\rangle_{x_{i+1}\in \mathcal T}$, which represents the model's confidence that each token being the next-token given on all previous tokens.
The Min-K\%~\cite{shi2023detecting} attack utilizes the fact that the model is less likely to predict words in the membership sentence with low probabilities.
Therefore, this attack selects K\% of tokens with the lowest predicted probabilities and calculates the score function as the average log likelihood of these selected tokens, i.e.,
\begin{equation}
\label{mink}
\mathcal{S}_\text{Min-K\%}(x;\mathcal M)=-\frac{\sum_{x_{i+1}\in \text{Min-K\%}(x)}\log p_{i,x_{i+1}}}{|\text{Min-K\%}(x)|}.
\end{equation}
The Min-K\%++~\cite{zhang2024min} attack further utilizes the probability of the nonmember tokens, resulting in the score function 
\begin{equation}
\label{minkpp}
\mathcal{S}_\text{Min-K\%++}(x;\mathcal M)=-\frac{\sum_{x_{i+1}\in \text{Min-K\%}(x)}\frac{\log p_{i,x_{i+1}}-\mu({\log\xl p_i})}{\sigma(\log\xl p_i)}}{|\text{Min-K\%}(x)|}.
\end{equation}
Another internal state is \textit{the gradients of the fine-tuned model with respect to the sample loss}, i.e., ${\partial \mathcal{L}}/{\partial \Delta}(x)$.
Because these gradients tend to be smaller for members, Wang et. al.~\cite{wang2024pandoras} proposed using the norm of these gradients as the score function, i.e., 
\begin{equation}
\label{gdt}
\mathcal{S}_{\text{GradNorm}_\theta}(x;\mathcal M)=-\Vert {\partial \mathcal{L}}/{\partial \Delta}(x) \Vert.
\end{equation}
They also suggested that \textit{the gradients on the input embeddings} (${\partial \mathcal{L}}/{\partial \xl e}$) could serve as an approximation of $ {\partial \mathcal{L}}/{\partial \Delta}$, resulting in the score function \begin{equation}
\label{gdx}
\mathcal{S}_{\text{GradNorm}_x}(x;\mathcal M)=-\Vert {\partial \mathcal{L}}/{\partial \xl e}(x) \Vert.
\end{equation}

\mypara{Augmented Perturbations}
In addition to performing standard forward and backward propagation, the adversary may introduce \textit{augmented perturbations} into the pipeline to collect internal states of nonmembers.
There are two approaches to introduce perturbations, as represented by the dashed line in~\Cref{fig:pipeline}.
The neighborhood attack~\cite{mattern-etal-2023-membership} perturbs the sample $x$ into $N$ paraphrased samples $\tilde{x}_1, \dots, \tilde{x}_N$ by replacing random words with predicted ones using mask-filling models like BERT~\cite{devlin2019bertpretrainingdeepbidirectional} or T5~\cite{2020t5}.
The losses of these paraphrased samples are then collected to calculate the score function as \begin{equation}
\label{nei}
\mathcal{S}_{\text{Nei}}(x;\mathcal M)=\frac{1}{N}\sum^N_{i=1} \mathcal L(\tilde{x}_i;\mathcal M)-\mathcal L(x;\mathcal M).
\end{equation}
The MoPe attack~\cite{mattern-etal-2023-membership} adds Gaussian noise to the model parameters, resulting $N$ perturbed models $\tilde{\mathcal M}_1,\dots, \tilde{\mathcal M}_N$.
The sample's loss on the perturbed models are then collected to calculate the score function as \begin{equation}
\label{gd}
\mathcal{S}_{\text{MoPe}}(x;\mathcal M)=\frac{1}{N}\sum^N_{i=1} \mathcal L(x;\tilde{\mathcal M}_i)-\mathcal L(x;\mathcal M).
\end{equation}

\mypara{Referenced Calibrations}
In addition to using internal states for membership inference, the adversary can employ external references to calibrate the score function. 
For instance, Carlini et al.~\cite{carlini2021extracting} proposed calibrating the sample loss with the entropy of input data evaluated by the zlib compressor, i.e., 
\begin{equation}
\label{zlib}
\mathcal{S}_{\text{zlib}}(x;\mathcal M)=|\text{zlib}(x)|-S_{LOSS}(x;\mathcal M).
\end{equation}
Additionally, other models can serve as effective references for calibration.
For example, the LiRA attack~\cite{mireshghallah-etal-2022-quantifying} utilizes the pre-trained model as a reference to calibrate the fine-tuned model's LOSS score function, i.e., 
\begin{equation}
\label{lira}
\mathcal{S}_{\text{LiRA}}(x;\mathcal M)=S_{LOSS}(x;\mathcal M_{pt})-S_{LOSS}(x;\mathcal M).
\end{equation}
The SPV-MIA attack~\cite{fu2023practical} trains a self-prompted model $\mathcal M_{sp}$ to calibrate the neighborhood score function, i.e., \begin{equation}
\label{spv}
\mathcal{S}_{\text{SPV}}(x;\mathcal M)=S_{Nei}(x;\mathcal M_{sp})-S_{Nei}(x;\mathcal M).
\end{equation}

\subsection{Pre-trained Model Calibration}
As highlighted in~\Cref{tab:meth}, while LoRA has emerged as the most widely adopted fine-tuning algorithm for LMs, only a limited number of existing MIAs have been explored against LoRA fine-tuning datasets.
Moreover, most attacks do not calibrate their scores with external references.
Given that the pre-trained model of the target LoRA model is publicly accessible, we propose \textit{pre-trained model calibration} to the performance of existing MIAs without incurring additional costs.
This technique leverages the pre-trained model as a reference to recalibrate the MIA scores.

Formally, let $\mathcal{S}$ denote a score function of an MIA targeting the LoRA fine-tuned model $\mathcal{M}$.
We introduce a calibrated score function, which utilizes the pre-trained model $\mathcal{M}{pt}$ as a reference, defined as: \begin{equation}
\label{enh}
\mathcal{S}_{\text{pt-ref}}(x;\mathcal{M})=\mathcal{S}(x;\mathcal{M}_{pt})-\mathcal{S}(x;\mathcal{M}).
\end{equation}
Here, $\mathcal{S}(x;\mathcal{M}_{pt})$ estimates $\Pr(x\in \mathcal{D}_{pt};\mathcal{M}_{pt})$, which serves as the \textit{a priori} probability of the membership status of $x$.
Conversely, $\mathcal{S}(x;\mathcal{M})$ estimates $\Pr(x\in \mathcal{D}_{ft};\mathcal{M})$.
Therefore, this score function captures the variation in confidence before and after fine-tuning.
Compared to the original score function derived solely from the fine-tuned model, the calibrated one can make the membership status of members and nonmembers more distinguishable.

However, among all existing MIAs, only the LOSS attack has been enhanced using this calibration technique, as demonstrated in \Cref{lira}.
Consequently, despite the scoring functions that already utilize reference calibration ($S_\text{zlib}$, $S_\text{SPV}$) and the score function incompatible to $\mathcal{M}_{pt}$ ($\text{GradNorm}_\theta$), we can enhance five MIAs by \Cref{enh}, including $S_\text{Min-k\%}, S_\text{Min-k\%++}, S_{\text{GradNorm}_x}, S_\text{Nei},$ and $S_\text{MoPe}$.

Ultimately, as shown in~\Cref{tab:meth}, our proposed \leak framework encompasses a total of fifteen MIA attacks, including five newly introduced MIAs.

\section{Evaluation}
In this section, we will evaluate all membership inference attacks in \leak.
First, we will outline our experimental settings and considerations for membership inference evaluation.
Next, we will analyze their applicability to fine-tuning datasets and demonstrate the superiority of our pre-trained model calibration method.
Finally, we will discuss the privacy risk of LoRA fine-tuning under different hyperparameters using \leak.

\subsection{Experimental Settings}
\mypara{Models}
We select three representative open-source LMs as our pre-trained models for fine-tuning, i.e., GPT-2 XL~\cite{radford2019language}, Pythia-2.8B~\cite{biderman2023pythia}, and Llama-2 7B~\cite{touvron2023llama}, with parameter sizes ranging from 1.5 billion, 2.8 billion, and 7 billion, respectively.
The chosen pre-trained models have been extensively used for LoRA fine-tuning.

\mypara{Datasets}
We focus on the following three downstream tasks to fine-tune LMs and evaluate the MIA risks.

\begin{itemize}
\item \textbf{AG News}~\cite{NIPS2015_250cf8b5} is a text classification task that categorizes news articles into four classes based on their titles and contents.
We leverage this dataset to simulate scenarios where LLMs are fine-tuned to adherent to specific output formats.

\item \textbf{Open Assistant Conversations (OAsst)}~\cite{10.5555/3666122.3668186} is a textual dataset containing multi-round conversations between users and AI chatbots in real-world scenarios.
In our evaluation, we utilize its subset, the OpenAssistant TOP-1 Conversation Threads dataset~\cite{oasst}, which is particularly suitable for fine-tuning LLMs into general-purpose chatbots.
The conversations of this dataset are converted to ChatML format~\cite{ChatML}.

\item \textbf{MedQA}~\cite{jin2020disease} is an English single-choice Question Answering (QA) task for medical exams.
Each question has five options to choose from.
We leverage MedQA to mimic the scenarios where LLMs are fine-tuned on sensitive datasets for domain adaptation.
\end{itemize}

For the AG News and OAsst tasks, we randomly select 10,000 items to construct the fine-tuning dataset $\mathcal{D}_{ft}^{\rm AG}$ and $\mathcal{D}_{ft}^{\rm OA}$, respectively.
Similarly, for the MedQA task, we randomly select 8,000 samples to build $\mathcal{D}_{ft}^{\rm Med}$.
Additionally, for each of these three tasks, we randomly select 1,000 samples distinct from their respective $\mathcal{D}_{ft}$ to form their validation datasets $\mathcal{D}_{val}$ (e.g., $\mathcal{D}_{val}^{\rm AG}$).

To evaluate the effectiveness of MIAs, for each task, we randomly select 512 items from its $\mathcal{D}_{ft}$ as the members to infer. 
Meanwhile, we randomly select 512 items that are disjoint from its $\mathcal{D}_{ft} \cup \mathcal{D}_{val}$ as non-members.

\mypara{Metrics}
We focus on three key metrics: PPL@val, GAP, and AUC.
These metrics assess the performance of the fine-tuned models, the susceptibility of fine-tuned models to MIAs, and the effectiveness of MIAs, respectively.

\begin{itemize}

\item \textbf{Model Utility:}
Perplexity (PPL) reflects the model's uncertainty regarding a given text.
We leverage the model's PPL on the \textit{validation} set (denoted as PPL@val) as an indicator of its specialized utility.
PPL can be calculated through~\Cref{ppl}.
A lower PPL@val suggests that the model exhibits better utility. 

\item \textbf{Overfitting Level:}
The generalization gap (GAP) is defined as the perplexity difference between the fine-tuning dataset and the validation dataset, i.e.,
\begin{equation}
\text{GAP} = \text{PPL@val} - \text{PPL@ft}.
\end{equation}
Since membership inference threat is primarily due to overfitting~\cite{8429311}, we use this metric to evaluate the overfitting level of the fine-tuned models on the fine-tuning datasets, thereby reflecting their hardnesses against MIAs.
A lower GAP value (around zero) implies the model is less overfitting, making membership inference more challenging and practical.
\item \textbf{Effectiveness of MIAs:}
We use the Area Under the Receiver Operating Characteristic Curve (AUC) to evaluate the effectiveness of MIAs.
A higher AUC indicates that an MIA is more effective at distinguishing between members and non-members.
The reason to use AUC lies in the fact that all attacks in \leak predict a score of membership rather than making a hard decision.
Therefore, varying thresholds yield the dynamic change of false-positive rates and true-positive rates, and AUC could capture this characteristic.
\end{itemize}

\begin{figure*}[t]
\centering
\includegraphics[width=0.92\linewidth]{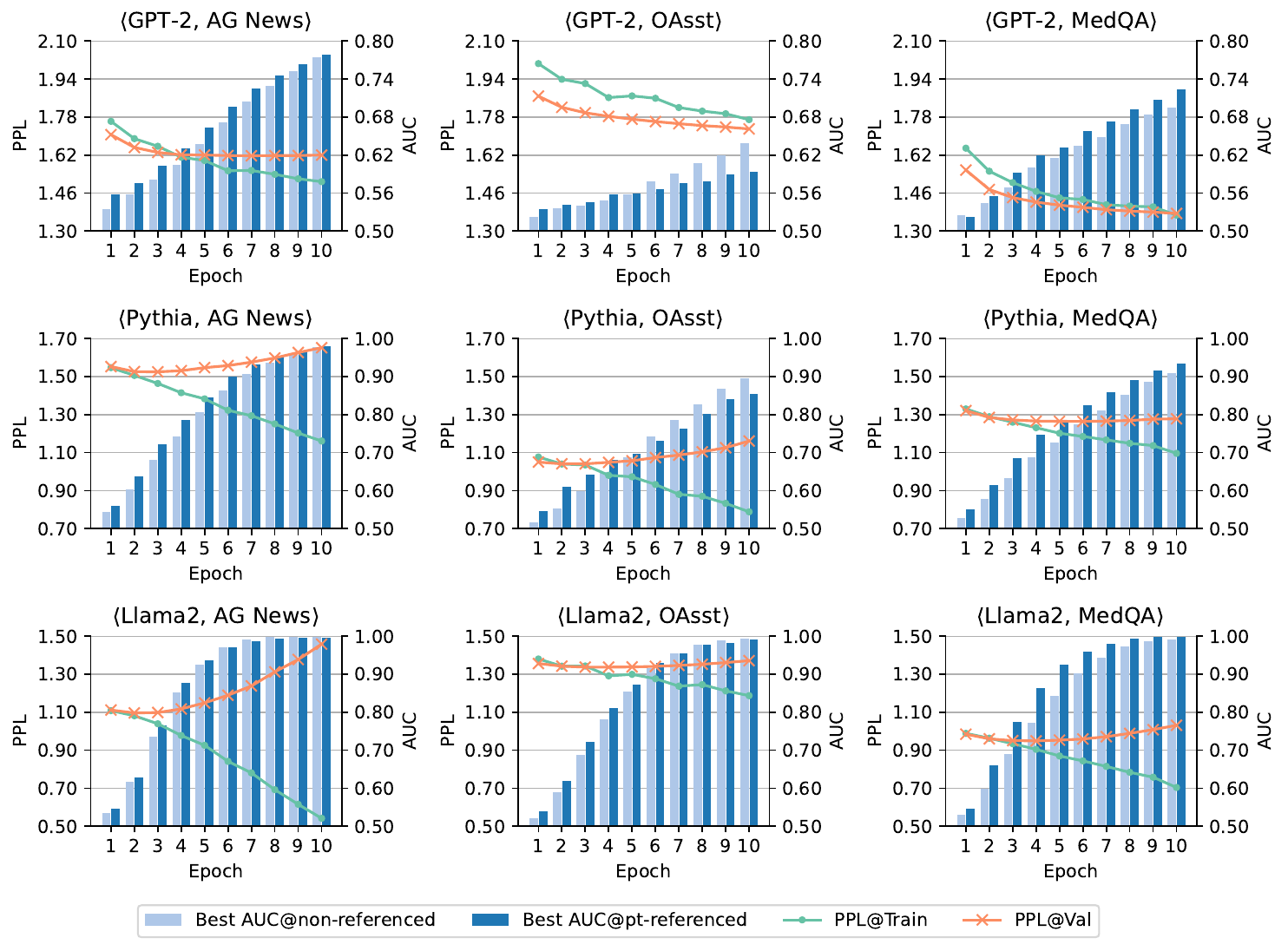}
\caption{
The perplexity of training and validation dataset as the fine-tuning epoch increases, as well as the best AUC achieved for non-referenced and pt-referenced MIAs.}
\label{fig:epoch}
\end{figure*}

\begin{table}[t]
\caption{The AUC of different MIAs against models fine-tuned from Llama-2. 
We highlight the results of the \colorbox{blue!10}{pt-referenced attacks}.}
\label{tab:auc_llama}
\centering
\begin{tabular}{c|ccc}
\toprule
\textbf{Attacks} & \textbf{AG News} & \textbf{OAsst} & \textbf{MedQA} \\
\midrule \midrule
zlib & 0.640 & 0.530 & 0.575 \\
\midrule
GradNorm$_\theta$ & 0.669 & 0.546 & 0.600 \\
\midrule
LOSS & 0.648 & 0.530 & 0.600 \\
\cellcolor{blue!10}{~~~~~~~~~~$\hookrightarrow$+\textit{Pre}} & \cellcolor{blue!10}{0.705} & \cellcolor{blue!10}{0.583} & \cellcolor{blue!10}{0.609} \\
\midrule
Neighborhood & 0.613 & 0.500 & 0.551 \\
\cellcolor{blue!10}{~~~~~~~~~~$\hookrightarrow$+\textit{Pre}} & \cellcolor{blue!10}{0.718} & \cellcolor{blue!10}{0.646} & \cellcolor{blue!10}{0.646} \\
\midrule
Min-K\% & 0.664 & 0.550 & 0.635 \\
\cellcolor{blue!10}{~~~~~~~~~~$\hookrightarrow$+\textit{Pre}} & \cellcolor{blue!10}{0.731} & \cellcolor{blue!10}{0.595} & \cellcolor{blue!10}{0.674} \\
\midrule
Min-K\%++ & 0.735 & 0.687 & 0.689 \\
\cellcolor{blue!10}{~~~~~~~~~~$\hookrightarrow$+\textit{Pre}} & \cellcolor{blue!10}{0.765} & \cellcolor{blue!10}{0.721} & \cellcolor{blue!10}{0.775} \\
\midrule
MoPe & 0.635 & 0.508 & 0.560 \\
\cellcolor{blue!10}{~~~~~~~~~~$\hookrightarrow$+\textit{Pre}} & \cellcolor{blue!10}{0.731} & \cellcolor{blue!10}{0.561} & \cellcolor{blue!10}{0.640} \\
\midrule
GradNorm$_x$ & 0.650 & 0.567 & 0.624 \\
\cellcolor{blue!10}{~~~~~~~~~~$\hookrightarrow$+\textit{Pre}} & \cellcolor{blue!10}{0.679} & \cellcolor{blue!10}{0.588} & \cellcolor{blue!10}{0.613} \\
\bottomrule
\end{tabular}
\end{table}

\subsection{Effectiveness of LoRA-Leak}
\label{sec:eval-main}

\mypara{Setups} 
In this section, we fine-tune all pre-trained models for 10 epochs.
For each epoch, we record the perplexity on the fine-tuning and validation sets to monitor the extent of overfitting.
Additionally, we perform \leak within each epoch and report the best AUC among the eight non-referenced MIAs and six referenced MIAs.
The experimental results are shown in~\Cref{fig:epoch}.

\mypara{Utility of Target Models}
According to \Cref{fig:epoch}, we could observe that LoRA fine-tuning effectively reduces the perplexity on training samples, indicating that it helps the model memorize information about the training data.
However, as the number of epochs increases, the perplexity on validation samples first decreases and then starts to increase.
Consequently, the model’s performance on downstream tasks initially improves and then declines.

\mypara{Relationship of Overfitting Level to MIA Risks}
We notice that the perplexity gap between the fine-tuning set and the validation set increases as fine-tuning progresses, indicating that the model’s overfitting level intensifies.
Consequently, the effectiveness of all MIAs also increase with this growing gap.
This observation highlights the significant impact of overfitting on the risk of membership inference.
Notably, the best AUC can approach 1.0 when the model is trained for 10 epochs, especially for Llama-2.
However, the advantage of non-referenced MIAs versus pre-trained model-referenced MIAs varies.
For instance, among all models fine-tuned on MedQA, the pre-trained model-referenced MIA consistently outperforms the non-referenced MIA.
Conversely, for models fine-tuned on OAsst, the pre-trained model-referenced MIA performs well when the model has not been severely overfitted, but the non-referenced MIA gains an advantage as overfitting intensifies.
This is because the scale of the membership score shrinks as fine-tuning progresses, making the calibrated score less applicable.

\mypara{Practical Considerations for Evaluating \leak}
As illustrated above, higher epochs can lead to overfitting, enhancing the effectiveness of MIAs and achieving higher reported metrics.
However, in practical scenarios, the model will be fine-tuned for optimal performance on downstream tasks rather than for the lowest perplexity on the training set.
Moreover, as shown in \Cref{fig:each-epochs} of the Supplementary Material, all attacks will achieve similarly high AUCs as overfitting intensifies, making their effectiveness indistinguishable.
Therefore, we believe that the setting of 3 epochs, where the validation perplexity is relatively low, is a fair representation of the real-world risk of LoRA fine-tuning.

\begin{figure*}[t]
\centering
\includegraphics[width=1\linewidth]{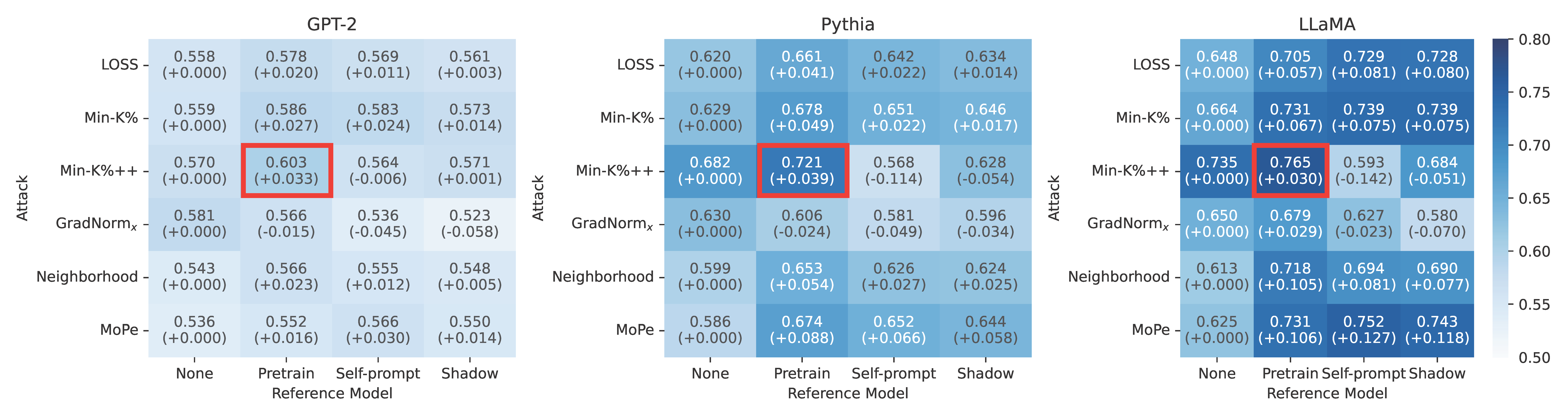}
\caption{The AUC achieved for three models trained on AG News, using different reference models to enhance the non-referenced MIAs. We box the highest AUC value in red.}
\label{fig:ref-model-discuss}
\end{figure*}

\mypara{Non-referenced MIAs' Performance}
We first perform the non-referenced MIAs as baselines.
Here we fix the epoch number to 3.
The AUC for these MIAs aginst Llama-2 are reported in \Cref{tab:auc_llama}, and the AUC achieved for other models are reported in \Cref{tab:main} of the Supplementary Material.
Our findings reveal that the attacks utilize more information could infer membership better, such as logits for Min-K and Min-K\%++ and gradients for $\text{GradNorm}_{\theta}$ and $\text{GradNorm}_{x}$.
Specifically, Min-K\%++ consistently acts as the most effective attack for models fine-tuned from Pythia and Llama-2.
However, for GPT-2, Min-K\%++ only outperforms other attacks on the MedQA dataset, while $\text{GradNorm}_{x}$ is more effective on the AG News and OAsst datasets.
Notably, some attacks perform worse than the LOSS attack, including the zlib attack, Neighborhood attack, and MoPe attack, even though they perform well on inferring membership of pre-training data~\cite{carlini2021extracting, mattern-etal-2023-membership, li-etal-2023-mope}.
We hypothesize that this discrepancy is due to the specific characteristics of fine-tuning data and LoRA modules:
(1) Fine-tuning data often involves domain-specific knowledge rather than general corpus data.
Consequently, it tends to have high entropy and hard to paraphrase, affecting the effectiveness of zlib and Neighborhood attacks.
(2) The compactness of LoRA parameters makes them sensitive to perturbations, potentially impacting the effectiveness of MoPe attacks.

\mypara{Pt-referenced MIAs' Performance}
We further perform pt-referenced MIAs against all nine target fine-tuned LMs. 
As~\Cref{tab:auc_llama} and~\Cref{tab:main} shows, we could observe that using pre-trained models as references can enhance the performance of each attack compared with the corresponding baselines in most cases.
For instance, introducing pre-trained Llama-2 further enhances the Min-K\%++ attack from 0.689 to 0.775 on MedQA.
Additionally, even though MoPe is not the best attack among non-referenced MIAs, its pt-referenced variant performs as the best attack against GPT-2 and Pythia on OAsst dataset.

\begin{tcolorbox}[colback=black!3!white,enhanced jigsaw,breakable]
\textbf{\textit{Takeaways: }}
Using the corresponding pre-trained model as a reference can amplify the effectiveness of existing MIAs and serve as a more powerful tool for privacy auditing in the context of LoRA fine-tuning.
\end{tcolorbox}

\subsection{Impact of Different Reference Models}
\label{sec:ref-models}

Despite using pre-trained models as a reference, some MIAs utilize other models for comparison. For instance, the LiRA attack adjusts the loss signal by comparing it to the loss of a shadow model fine-tuned on a dataset with similar distributions~\cite{9833649}. Additionally, Fu et al.~\cite{9833649} propose constructing a shadow model by prompting the target model itself. In this section, we aim to assess the effectiveness of these different reference models.

\mypara{Setups}
We first construct shadow models following \cite{9833649} by fine-tuning three pre-trained models on the TLDR News dataset~\cite{JulesBelveze_tldr_news}, which shares similar domains with AG News.
Additionally, we create self-prompt models following~\cite{fu2023practical} by using the first 16 words of the TLDR News dataset to prompt the target model, followed by fine-tuning three pre-trained models on the resulting corpus.
We fix the downstream task as AG News.
The AUC results are presented in \Cref{fig:ref-model-discuss}.

\mypara{Results}
Our evaluation reveals that both shadow models and self-prompt models enhance the claimed baseline attacks.
However, they are generally less effective than using the pre-trained model as a reference.
For example, across all attacks, using the pre-trained model as the reference could enhance Min-K\%++-$\text{Ref}_\text{pt}$ and achieve the most effective MIA among all attacks while using another model as the reference even degrades its AUC.
Moreover, we observe that some attacks, such as the Min-K\% attack, Neighbourhood attack, and MoPe attack can be boosted by any reference models, resulting in AUC increases of 0.067 to 0.127 for Llama-2.
Furthermore, other reference models require additional datasets and fine-tuning efforts, while the pre-trained model is naturally obtainable within the context of LoRA fine-tuning.

\begin{tcolorbox}[colback=black!3!white,enhanced jigsaw,breakable]
\textbf{\textit{Takeaways: }}The pre-trained model is the most effective reference model with the most convenience in the context of LoRA fine-tuning.
\end{tcolorbox}

\subsection{Impact of Fine-tuning Modules}
\label{sec:module}

\begin{table}[t]
\centering
\caption{
Discussion on fine-tuning modules. Here we report the MIA AUC results (Non-referenced$\rightarrow$pt-referenced). 
We highlight \colorbox{blue!10}{the three lowest AUC results}. 
The fine-tuning dataset is AG News.
}
\label{tab:module}
\setlength{\tabcolsep}{4pt}
\begin{tabular}{c|l|rr|c}
\toprule
Model & Fine-tuning Module & $r$ & $\alpha$  & MIA AUC \\ \midrule \midrule
\multirow{8}{*}{GPT-2}&qkv, o, u, d & 4 & 8 & 0.581$\rightarrow$\textbf{0.603}
\\
&qkv, o, u & 5 & 10  & 0.587$\rightarrow$\textbf{0.608}
 \\
&\cellcolor{blue!10}{qkv, o,~~~~~d} & \cellcolor{blue!10}{5} & \cellcolor{blue!10}{10} & \cellcolor{blue!10}{0.533$\rightarrow$\textbf{0.566}}
 \\
&~~~~~~~~~~~~u, d & 6 & 12 & 0.579$\rightarrow$\textbf{0.614}
 \\
&\cellcolor{blue!10}{qkv, o}& \cellcolor{blue!10}{10} & \cellcolor{blue!10}{20} & \cellcolor{blue!10}{0.536$\rightarrow$\textbf{0.572}}
 \\
&~~~~~~~~~~~~u & 12 & 24  & 0.595$\rightarrow$\textbf{0.626}
 \\
&\cellcolor{blue!10}{~~~~~~~~~~~~~~~~d} & \cellcolor{blue!10}{12} & \cellcolor{blue!10}{24} & \cellcolor{blue!10}{0.530$\rightarrow$\textbf{0.559}}
 \\
&qkv & 16 & 32 & 0.545$\rightarrow$\textbf{0.573}
 \\ \midrule
\multirow{8}{*}{Pythia}&qkv, o, u, d & 4 & 8 & 0.682$\rightarrow$\textbf{0.721}
 \\
&qkv, o, u & 5 & 10 & 0.682$\rightarrow$\textbf{0.723}
 \\
&\cellcolor{blue!10}{qkv, o,~~~~~d} & \cellcolor{blue!10}{5} & \cellcolor{blue!10}{10} & \cellcolor{blue!10}{0.574$\rightarrow$\textbf{0.601}}
 \\
&~~~~~~~~~~~~u, d & 6 & 12 & 0.676$\rightarrow$\textbf{0.717}
 \\
&\cellcolor{blue!10}{qkv, o}& \cellcolor{blue!10}{10} & \cellcolor{blue!10}{20} & \cellcolor{blue!10}{0.576$\rightarrow$\textbf{0.609}}
 \\
&~~~~~~~~~~~~u & 12 & 24 & 0.685$\rightarrow$\textbf{0.728}
 \\
&\cellcolor{blue!10}{~~~~~~~~~~~~~~~~d} & \cellcolor{blue!10}{12} & \cellcolor{blue!10}{24} & \cellcolor{blue!10}{0.553$\rightarrow$\textbf{0.584}}
 \\
&qkv & 16 & 32 & 0.579$\rightarrow$\textbf{0.609}
 \\ \midrule
\multirow{12}{*}{Llama-2}&qkv, o, u, d, g & 4 & 8 & 0.735$\rightarrow$\textbf{0.765}
 \\
&qkv, o, u,~~~~~g& 4 & 8 & 0.675$\rightarrow$\textbf{0.702}
 \\
&qkv, o, u, d & 4 & 8 & 0.659$\rightarrow$\textbf{0.685} \\
&qkv, o,~~~~~d, g & 4 & 8 & 0.681$\rightarrow$\textbf{0.706}
 \\
&qkv, o,~~~~~d & 6 & 12 & 0.603$\rightarrow$\textbf{0.620}
 \\
&~~~~~~~~~~~ u, d, g & 6 & 12 & 0.727$\rightarrow$\textbf{0.755}
 \\
&\cellcolor{blue!10}{qkv, o}& \cellcolor{blue!10}{8} & \cellcolor{blue!10}{16} & \cellcolor{blue!10}{0.588$\rightarrow$\textbf{0.601}}
 \\
&~~~~~~~~~~~ u,~~~~ g & 9 & 18 & 0.710$\rightarrow$\textbf{0.739}
 \\
&\cellcolor{blue!10}{qkv}& \cellcolor{blue!10}{11} & \cellcolor{blue!10}{22} & \cellcolor{blue!10}{0.583$\rightarrow$\textbf{0.597}}\\
&~~~~~~~~~~~ u & 18 & 36 & 0.664$\rightarrow$\textbf{0.692}
 \\
&~~~~~~~~~~~~~~~~~~~ g & 18 & 36 & 0.680$\rightarrow$\textbf{0.706}
 \\
& \cellcolor{blue!10}{~~~~~~~~~~~~~~~ d} & \cellcolor{blue!10}{18} & \cellcolor{blue!10}{36} & \cellcolor{blue!10}{0.574$\rightarrow$\textbf{0.585}}
 \\ \bottomrule
\end{tabular}
\end{table}

\mypara{Setups}
Considering that LoRA adapters can be added to different modules of LMs, in this part, we aim to discuss the impact of different fine-tuning choices on the effectiveness of MIAs.
We fix the fine-tuning dataset to AG News and keep all other hyperparameters the same as in~\Cref{sec:default-train}.

\mypara{Results}
As shown in \Cref{tab:module}, we could observe that fine-tuning different modules results in varying susceptibility to MIAs.
Specifically, excluding \texttt{attention layers} (qkv) and \texttt{downscale layers} (d) from the feed-forward layers has minimal impact on the AUC of the best \leak attacks.
However, excluding the \texttt{upscale layer} (u) during fine-tuning could significantly reduce the best MIA AUC.
For instance, for all three models, the fine-tuning modules corresponding to the three lowest MIA AUC experimental results did not involve the \texttt{upscale layer}.
Meanwhile, we also evaluate whether excluding certain modules during the fine-tuning process would affect model performance.
When we only fine-tune the \texttt{downscale} \texttt{layers}, the results of PPL@val are 1.680, 1.543, and 1.110 for GPT-2, Pythia, and Llama-2, respectively, while the results will be 1.631, 1.525, and 1.098 when we include all layers.
This means the impact on model performance is relatively low when excluding certain modules.
Notably, regardless of whether specific modules are excluded, incorporating information from pre-trained models consistently enhances the effectiveness of MIAs.

\begin{tcolorbox}[colback=black!3!white,enhanced jigsaw,breakable]
\textbf{\textit{Takeaways: }}Fine-tuning different modules results in varying susceptibility to MIAs.
Specifically, including the upscale layers (u) during LoRA fine-tuning makes the model more vulnerable to MIAs.
\end{tcolorbox}

\section{Defenses}

In this section, we discuss several potential defense mechanisms against \leak.
We first discuss three typical fine-tuning techniques that could potentially defend against MIAs, including \textit{dropout}, \textit{weight decay}, and \textit{differential privacy}.
Additionally, based on our findings from~\Cref{sec:module}, we further explore a novel defense approach that excludes vulnerable layers during LoRA fine-tuning.

Given that Llama-2 exhibits the highest susceptibility to membership inference and has become the most widely deployed pre-trained model in the real world, we focus on Llama-2 throughout this section.
For ease of discussion, we explore the effectiveness of each method independently, while keeping other irrelevant settings consistent with those in~\Cref{sec:default-train}.

\subsection{Dropout}
\label{sec:def-dropout}

\begin{table*}[t]
\caption{
Results on weight decay defense.
We report the best MIA AUC results (non-referenced$\rightarrow$pt-referenced) with varying weight decay rates ($\lambda$).
}
\label{tab:defense-decay}
\centering

\begin{tabular}{c|crc|crc|crc}
\toprule
 \multirow{2}{*}{$\lambda$} & \multicolumn{3}{c|}{AG News} & \multicolumn{3}{c|}{OAsst} & \multicolumn{3}{c}{MedQA} \\\cmidrule(lr){2-4}  \cmidrule(lr){5-7} \cmidrule(lr){8-10}
~ & \multicolumn{1}{c}{PPL@val} & \multicolumn{1}{c}{GAP} &\multicolumn{1}{c|}{Best AUC}& \multicolumn{1}{c}{PPL@val} & \multicolumn{1}{c}{GAP} &\multicolumn{1}{c|}{Best AUC}& \multicolumn{1}{c}{PPL@val} & \multicolumn{1}{c}{GAP}&\multicolumn{1}{c}{Best AUC} \\ \midrule\midrule
w/o & 1.098 & 0.059 & 0.735$\rightarrow$\textbf{0.765} & 1.041 & 0.007 & 0.687$\rightarrow$\textbf{0.721} & 0.950 & 0.021 & 0.689$\rightarrow$\textbf{0.775} \\ \midrule
$10^{-4}$ & 1.099 & 0.060 & 0.736$\rightarrow$\textbf{0.764} & 1.043 & 0.008 & 0.687$\rightarrow$\textbf{0.722} & 0.953 & 0.023 & 0.689$\rightarrow$\textbf{0.778} \\
$10^{-3}$ & 1.098 & 0.058 & 0.733$\rightarrow$\textbf{0.764} & 1.043 & 0.008 & 0.685$\rightarrow$\textbf{0.721} & 0.952 & 0.024 & 0.688$\rightarrow$\textbf{0.776} \\
$10^{-2}$ & 1.099 & 0.059 & 0.736$\rightarrow$\textbf{0.763} & 1.042 & 0.008 & 0.688$\rightarrow$\textbf{0.723} & 0.952 & 0.023 & 0.689$\rightarrow$\textbf{0.775} \\
$10^{-1}$ & 1.098 & 0.059 & 0.737$\rightarrow$\textbf{0.766} & 1.043 & 0.008 & 0.687$\rightarrow$\textbf{0.721} & 0.953 & 0.024 & 0.690$\rightarrow$\textbf{0.777} \\
\bottomrule
\end{tabular}
\end{table*}

\begin{table*}[t]
\caption{
Results on differential privacy defense.
We report the best MIA AUC results (non-referenced$\rightarrow$pt-referenced) with varying privacy budgets ($\epsilon$).
}
\label{tab:defense-dp}
\centering
\begin{tabular}{c|crc|crc|crc}
\toprule
 \multirow{2}{*}{$\epsilon$} & \multicolumn{3}{c|}{AG News} & \multicolumn{3}{c|}{OAsst} & \multicolumn{3}{c}{MedQA} \\\cmidrule(lr){2-4}  \cmidrule(lr){5-7} \cmidrule(lr){8-10}
~ & \multicolumn{1}{c}{PPL@val} & \multicolumn{1}{c}{GAP} &\multicolumn{1}{c|}{Best AUC}& \multicolumn{1}{c}{PPL@val} & \multicolumn{1}{c}{GAP} &\multicolumn{1}{c|}{Best AUC}& \multicolumn{1}{c}{PPL@val} & \multicolumn{1}{c}{GAP}&\multicolumn{1}{c}{Best AUC} \\ \midrule\midrule
w/o & 1.098 & 0.059 & 0.735$\rightarrow$\textbf{0.765} & 1.041 & 0.007 & 0.687$\rightarrow$\textbf{0.721} & 0.950 & 0.021 & 0.689$\rightarrow$\textbf{0.775} \\ \midrule
0.1 & 1.969 & -0.031 & 0.521$\rightarrow$\textbf{0.531} & 1.255 & -0.054 & 0.508$\rightarrow$\textbf{0.510} & 1.473 & -0.004 & \textbf{0.516}$\rightarrow$\textbf{0.516} \\
1.0 & 1.256 & -0.015 & 0.519$\rightarrow$\textbf{0.527} & 1.093 & -0.039 & \textbf{0.520}$\rightarrow$0.506 & 1.134 & -0.010 & \textbf{0.515}$\rightarrow$0.506 \\
10 & 1.223 & -0.011 & 0.521$\rightarrow$\textbf{0.539} & 1.080 & -0.034 & 0.524$\rightarrow$\textbf{0.525} & 1.090 & -0.009 & \textbf{0.527}$\rightarrow$0.520 \\
\bottomrule
\end{tabular}
\end{table*}

\mypara{Definition}
Dropout~\cite{JMLR:v15:srivastava14a} is a traditional technique used to mitigate overfitting in deep learning models.
It works by randomly deactivating partial neurons and their corresponding connections during each training step.
By doing so, the dependence between neurons is reduced, which in turn decreases the risk of MIAs.
We leverage the dropout rate, denoted as $\eta \in [0,1]$, to determine the proportion of trainable neurons to be dropped.
The experimental results are shown in~\Cref{tab:defense-dropout}.
The AUC results for each specific MIA are shown in~\Cref{fig:each-dropout}.

\mypara{Results}
We first could observe that the fine-tuned models become less vulnerable to MIAs as the dropout rate $\eta$ increases.
Simultaneously, the overall performance of the models remains relatively stable even at $\eta=0.95$, where the best MIA AUC could achieve a reduction of at most 0.154 (i.e., on the AG News dataset).
One extreme case is that, when $\eta=0.99$, the AUC of the best MIA on the OAsst dataset could even decrease to 0.543, however, the model utility also degrades. 
Considering that dropout also incurs negligible overhead in terms of computational cost, it is advisable to incorporate dropout with the LoRA fine-tuning process.
Nevertheless, it’s worth noting that introducing the pre-trained model as a reference could still result in a stronger attack even when leveraging dropout.

\begin{tcolorbox}[colback=black!3!white,enhanced jigsaw,breakable]
\textbf{\textit{Takeaways: }}Combining dropout with LoRA can mitigate the risk of membership inference, especially in the context of a high dropout rate. 
Such mitigation will not compromise model utility significantly as long as the dropout rate remains within a reasonable bound.
\end{tcolorbox}

\begin{table*}[t]
\caption{
Results on excluding vulnerable layers-based defense.
We report the best MIA AUC results (non-referenced$\rightarrow$pt-referenced) with excluding varying layers.
}
\label{tab:defense-module}
\centering
\setlength{\tabcolsep}{1mm}{
\begin{tabular}{c|rr|crc|crc|crc}
\toprule
 \multirow{2}{*}{Target Module} & \multirow{2}{*}{$r$}& \multirow{2}{*}{$\alpha$}&\multicolumn{3}{c|}{AG News} & \multicolumn{3}{c|}{OAsst} & \multicolumn{3}{c}{MedQA} \\\cmidrule(lr){4-6}  \cmidrule(lr){7-9} \cmidrule(lr){10-12}
~ &&& \multicolumn{1}{c}{PPL@val} & \multicolumn{1}{c}{GAP} &\multicolumn{1}{c|}{Best AUC}& \multicolumn{1}{c}{PPL@val} & \multicolumn{1}{c}{GAP} &\multicolumn{1}{c|}{Best AUC}& \multicolumn{1}{c}{PPL@val} & \multicolumn{1}{c}{GAP}&\multicolumn{1}{c}{Best AUC} \\ \midrule\midrule
qkv, o, u, g, d & 4 & 8  & 1.099 & 0.059 & {0.735  $\rightarrow$\textbf{0.765}} & 1.043 & 0.009 & 0.687$\rightarrow$\textbf{0.721} & 0.952 & 0.024 & 0.689  $\rightarrow$\textbf{0.775} \\
qkv, o, u,\quad \enspace d & 4 & 8  & 1.098 & 0.037 & {0.659$\rightarrow$\textbf{0.685}} & 1.043 & -0.003 & 0.632$\rightarrow$\textbf{0.672} & 0.958 & 0.013 & {0.633$\rightarrow$\textbf{0.710}} \\
qkv, o,\quad \enspace g, d & 4 & 8  & 1.097 & 0.041 & {0.681$\rightarrow$\textbf{0.706}} & 1.043 & -0.001 & 0.645$\rightarrow$\textbf{0.679} & 0.957 & 0.017 & {0.652$\rightarrow$\textbf{0.734}} \\
qkv, o, \qquad \;d & 6 & 12  & 1.098 & 0.020 & {0.603$\rightarrow$\textbf{0.620}} & 1.045 & -0.016 & 0.595$\rightarrow$\textbf{0.628} & 0.969 & 0.003 &{0.592$\rightarrow$\textbf{0.654}} \\
\bottomrule
\end{tabular}}
\end{table*}

\subsection{Weight Decay}
\label{sec:def-weightdecay}

\mypara{Definition}
Krogh et al.~\cite{krogh1991simple} propose that adding a penalty term for large weights to the original training loss function $\mathcal{L}$ can improve the generalization of machine learning models, thereby alleviating overfitting. 
Given that membership inference is primarily influenced by overfitting, we thus investigate the effectiveness of weight decay as a defense.
To this end, the modified loss function $\mathcal{L}_{decay}$ for the LoRA fine-tuning can be defined as
\begin{equation}
\mathcal{L}_{decay} = \mathcal{L} + \frac{\lambda}{2} \Vert \Delta \Vert^2,
\end{equation}
where $\Delta$ represents the weights of the LoRA modules, and $\lambda$ is a hyperparameter that represents the weight decay rate.

\mypara{Setups}
We employ the Adam optimizer with the decoupled weight decay (AdamW) proposed by Loshchilov et al.~\cite{loshchilov2018decoupled} to fine-tune Llama-2 on three datasets. 
During fine-tuning, we vary the weight decay rate $\lambda$ from $10^{-4}$ to $10^{-1}$, while also considering models without any weight decay. The results of the best MIA AUC and model utilities are summarized in~\Cref{tab:defense-decay}.
The AUC results of each MIA are shown in~\Cref{fig:each-weight}.

\mypara{Results}
Our results reveal that applying weight decay has no significant impact on the model in terms of both MIA risks and performance. 
For instance, all the best AUC fluctuates less than 0.010 compared to the model without any weight decay.
Additionally, the perplexity on the validation set increases by at most 0.003.

\begin{tcolorbox}[colback=black!3!white,enhanced jigsaw,breakable]
\textbf{\textit{Takeaways: }}In the context of LoRA fine-tuning, weight decay cannot work for mitigating membership inference risks.
This conclusion aligns with~\cite{kaya2020effectiveness}, which demonstrates that weight decay can even exacerbate the risk of MIAs for Convolutional Neural Networks (CNNs).
\end{tcolorbox}

\subsection{Differential Privacy (DP)}
\label{sec:def-dp}

\mypara{Definition}
Differential Privacy (DP)~\cite{10.1007/11681878_14} is a classical privacy protection technique that provides rigorous indistinguishability for a single entry.
Since DP offers provable protection in terms of dataset privacy, it has become a natural defense against MIAs.
Specifically, a randomized algorithm $\mathcal M$ with output space $O$ achieves $\epsilon$-differential privacy if
\begin{equation*}
\Pr[\mathcal M(\mathcal{D})\in O]\le e^\epsilon\Pr[\mathcal M(\mathcal{D}')\in O]+\delta
\end{equation*}
holds for any two adjacent databases $\mathcal{D}$ and $\mathcal{D}'$ that differ only at one entry.
The parameter $\epsilon$ represents the privacy budget.
A smaller $\epsilon$ indicates stronger privacy guarantees.

\mypara{Setups}
We employ DPLoRA~\cite{yu2022differentially} implemented by DP-Transformers~\cite{dp-transformers} to fine-tune Llama-2, varying the privacy budget $\epsilon$ from 0.1 to 10.
The results related to DP are summarized in~\Cref{tab:defense-dp}.
Additionally, the training and validation loss during the fine-tuning process is depicted in \Cref{fig:ft-dp}, and the AUC results of each MIA are shown in~\Cref{fig:each-priv-budget}.

\mypara{Results}
We conduct the analysis from two perspectives: the effectiveness of mitigating MIA and the impact on model performance.
Regarding the effectiveness of DP, we observe that introducing DP can significantly reduce the susceptibility to MIAs: across all three datasets, the best AUCs among all MIAs drop to $\sim$0.5 after applying DP, whereas the best AUCs are above 0.7 without DP.
As the privacy budget decreases, the AUC of MIAs slightly decreases.
However, we notice that a smaller privacy budget does not provide substantial gains of MIA protection compared to larger $\epsilon$.
For instance, on the AG News dataset, using DP with $\epsilon=0.1$ only decreases the best MIA AUC by 0.008.
Moreover, as~\Cref{fig:each-priv-budget} shows, non-referenced attacks occasionally outperform the pt-referenced attacks, but the differences remain marginal.
This behavior suggests that DP effectively renders all attacks akin to random guessing.
Furthermore, though DP is effective in relieving \leak, the enhanced privacy comes at a significant cost to model utility.
Even with a large privacy budget (e.g., $\epsilon=10$), models still experience noticeable performance degradation, which deteriorates further as $\epsilon$ decreases (see~\Cref{fig:ft-dp}).
Besides the performance hit, we observe that the loss converges more slowly after applying DP with $\epsilon=1.0$ and $\epsilon=10$, and it converges at an even slower rate when $\epsilon=0.1$.
Additionally, DP introduces substantial computational overhead, i.e., fine-tuning on the AG News, OAsst, and MedQA datasets with DP takes 31$\times$, 7$\times$, and 11$\times$ longer runtime, respectively, compared to fine-tuning without DP.

\begin{tcolorbox}[colback=black!3!white,enhanced jigsaw,breakable]
\textbf{Takeaways:} While DP achieves nearly perfect defense against \leak, its performance impact and computational cost make it impractical for real-world deployment.
\end{tcolorbox}

\subsection{Excluding Vulnerable layers}
\label{sec:def-exc}

In light of the findings from~\Cref{sec:module}, we identify that some modules are the key contributors to membership inference vulnerability, thus we regard excluding these vulnerable layers as a new defense strategy for LoRA fine-tuning.

\mypara{Setups}
Recall that fine-tuning the \texttt{up} (u) and \texttt{gate} (g) layers of Llama-2 can amplify the risks of MIAs.
To mitigate this, we fine-tune Llama-2 across all three datasets using all LoRA modules (i.e., qkv, o, u, g, d) excluding u, g, or both of them.
Note that we adjust their rank $r$ to ensure that the numbers of the tuned parameters remain roughly the same across all models.
The best MIA AUC results are summarized in~\Cref{tab:defense-module}.
The AUC results of each MIA when excluding certain layers are shown in~\Cref{fig:each-specific-layer}.

\mypara{Results}
First, we can observe that excluding just one of the vulnerable modules can only reduce the best MIA AUC by 0.041 to 0.080, though removing the \textit{gate} layers has a more pronounced impact compared to removing the \textit{up} layers.
When both vulnerable modules are excluded, the best AUC values decrease significantly by 0.121 to 0.145.
Compared to dropout defense, excluding one of the modules roughly corresponds to the impact of a dropout rate with $\eta=0.85$, and excluding both modules is similar to a dropout rate with $\eta=0.95$.
However, pt-referenced MIAs still consistently perform as the most effective attack.
Regarding model utility, excluding these layers has minimal impact when fine-tuning on the AG News and OAsst datasets.
However, there is a moderate performance downgrade when fine-tuning on MedQA.
This discrepancy may be due to the specific correlation between the ability of medical knowledge and these excluded modules.

\begin{tcolorbox}[colback=black!3!white,enhanced jigsaw,breakable]
\textbf{\textit{Takeaways: }}Excluding the vulnerable layers provides a practical defense against MIAs.
However, performance degradation may occur when model knowledge is correlated with the excluded modules.
\end{tcolorbox}

\section{Limitations}

\mypara{Closed-source Language Models}
Nowadays, closed-source large models such as ChatGPT have progressively made fine-tuning capabilities available through APIs for data uploads~\cite{openai_api}. 
However, the output information of closed-source LLMs is insufficient to initiate MIAs.
For instance, OpenAI API~\cite{openai_api} and HuggingFace Serverless Inference API~\cite{huggingface_api} do not offer access to the loss on inputs, which is crucial for executing the \textit{so-called} black-box MIAs~\cite{jagannatha2021membershipinferenceattacksusceptibility}.
Additionally, the fine-tuning algorithms used by closed-source models are not publicly available, and users cannot flexibly design hyperparameters. 
Therefore, our work focuses on attacking open-source language models. 
We leave the design of label-only membership inference attacks~\cite{CTCP21,LZ21} against language models as our future work.

\mypara{The Scale of The Target Models}
Our evaluation includes a diverse set of models ranging in parameter size from 125M to 13B.
While we acknowledge that experimenting with even larger language models would be an exciting opportunity to explore more advanced capabilities, such an endeavor would exceed our current technical resources.
Moreover, given the scope of our study, we believe that the additional insights gained from models with significantly larger parameter sizes would be marginal, particularly in relation to the increased computational and infrastructural demands they would impose.
Therefore, we chose to focus on models within this parameter range, which strikes an optimal balance between technical feasibility and the value of the insights.

\section{Conclusion}
In this paper, we propose \leak, a comprehensive evaluation framework for MIAs against LMs.
In \leak, we consider eight non-referenced MIAs and six pt-referenced MIAs, which provides a systematic quantification for membership leakage.
By conducting experiment on three practical datasets with three different pre-trained language models, we present the superiority of the pt-referenced MIA attacks, which achieves the best performance among existing MIAs.
Meanwhile, to demonstrate the generality of our insights, we further fine-tune the LMs with LoRA variants, and launch \leak against these fine-tuned models.
We find that DoRA will slightly increase of risk of MIA, while qLoRA could mitigate MIA with the degrade of performance.
Additionally, we discuss four defenses and find that excluding specific LM layers and dropout can mitigate privacy risks.
We hope our work can benefit the community by presenting comprehensive insights for auditing the privacy risks of LMs, and provide valuable insights for selecting the optimal setting of LoRA fine-tuning to mitigate the risk of MIA.

\newpage

\bibliographystyle{IEEEtran}
\bstctlcite{IEEEexample:BSTcontrol}
\bibliography{bib}

\clearpage

\appendix
\section*{Dataset Examples}
\label{sec:data-example}
\subsection{AG News}
\begin{tcolorbox}[colback=gray!25!white, size=title,breakable,boxsep=1mm,colframe=white,before={\vskip1mm}, after={\vskip0mm}]
Below is a news article. Please classify it under one of the following classes (World, Business, Sports, Sci/Tech).\\\\
\#\#\# Article: Bangladesh paralyzed by strikes Opposition activists have brought many towns and cities in Bangladesh to a halt, the day after 18 people died in explosions at a political rally.\\\\
\#\#\# Class: World
\end{tcolorbox}
\subsection{OAsst}
\begin{tcolorbox}[colback=gray!25!white, size=title,breakable,boxsep=1mm,colframe=white,before={\vskip1mm}, after={\vskip0mm}]
\textless\textbar im\_start\textbar \textgreater user\\
Tell me a knock-knock joke.\textless\textbar im\_end\textbar \textgreater\\
\textless\textbar im\_start\textbar \textgreater assistant\\
Knock knock!\textless\textbar im\_end\textbar \textgreater\\
\textless\textbar im\_start\textbar \textgreater user\\
Who's there?\textless\textbar im\_end\textbar \textgreater\\
\textless\textbar im\_start\textbar \textgreater assistant\\
Boo.\textless\textbar im\_end\textbar \textgreater
\end{tcolorbox}
\subsection{MedQA}
\begin{tcolorbox}[colback=gray!25!white, size=title,breakable,boxsep=1mm,colframe=white,before={\vskip1mm}, after={\vskip0mm}]
Please answer the letter of option truthfully.\\\\
\#\#\# Question: Which of the following compounds is most responsible for the maintenance of appropriate coronary blood flow?? \\\\
\#\#\# Options: `A': `Epinephrine', `B': `Norepinephrine', `C': `Histamine', `D': `Nitric oxide', `E': `VEGF'\\\\
\#\#\# Answer: D
\end{tcolorbox}
\section*{Default Fine-tuning Settings of LoRA}
\label{sec:default-train}
We leverage the Supervised Fine-Tuning (SFT) command provided by the Transformer Reinforcement Learning (TRL) library \cite{vonwerra2022trl} for LoRA fine-tuning.
The LoRA modules are added on all linear layers of the pre-trained model, including both the self-attention modules and the feed-forward modules of the transformer.
These modules are configured with a rank ($r$) of 4, and their scaling factor ($\alpha$) is set to be twice of $r$.
Each model is trained with a batch size of 16 and a dropout rate of 5\%.
The default fine-tuning epoch number is 3.
We employ the AdamW optimizer \cite{loshchilov2018decoupled}, with a fixed learning rate of $10^{-4}$ without using the weight decay technique.
The text sequences are truncated to a maximum of 1024 tokens, without resorting to sequence packing or input masking techniques.
All other hyper-parameters are set as the default values of the script.

\section*{Impact of LoRA Variants}
\label{sec:varlora}

\mypara{Setups}
Given that multiple variants of LoRA have been developed to date, their associated privacy threats also urgently require assessment.
We here explore two popular LoRA variants, i.e., Weight-Decomposed Low-Rank Adaptation (DoRA)~\cite{liu2024doraweightdecomposedlowrankadaptation} and qLoRA~\cite{dettmers2023qlora}.
Specifically, we consider the Int8 quantization and FP4 quantization settings for qLoRA.
The pre-trained model in this part is Llama-2.
We evaluate both the membership inference risks and the performance of these LoRA variants.


\mypara{Results}
As \Cref{tab:lora-methods} illustrates, using DoRA to fine-tune models slightly increases the effectiveness of \leak by 0.004 to 0.008.
However, the fine-tuned models experience a slight decrease in overall performance.
For qLoRA, the effectiveness of \leak decreases by -0.001 to 0.014 for Int8 quantization, and FP4 quantization leads to a further decrease of 0.024 to 0.030.
Additionally, qLoRA fine-tuning also exhibits performance degradation.
Therefore, we could conclude that lower-precision quantization for qLoRA reduces vulnerability but at the cost of greater performance degradation.

\begin{tcolorbox}[colback=black!3!white,enhanced jigsaw,breakable]
\textbf{\textit{Takeaways: }}Introducing pre-trained models can enhance the effectiveness of MIA attacks across different LoRA variants, indicating the presence of general privacy vulnerabilities within the LoRA paradigm.
\end{tcolorbox}

\begin{table*}[t]
\caption{
Results on LoRA variants.
We report the best MIA AUC (non-referenced$\rightarrow$pt-referenced).
The model is Llama-2.
}
\label{tab:lora-methods}
\centering
\setlength{\tabcolsep}{1.2mm}{
\begin{tabular}{c|crc|crc|crc}
\toprule
 \multirow{2}{*}{Method} & \multicolumn{3}{c|}{AG News} & \multicolumn{3}{c|}{OAsst} & \multicolumn{3}{c}{MedQA} \\\cmidrule(lr){2-4}  \cmidrule(lr){5-7} \cmidrule(lr){8-10}
~ & \multicolumn{1}{c}{PPL@val} & \multicolumn{1}{c}{GAP} &\multicolumn{1}{c|}{Best AUC}& \multicolumn{1}{c}{PPL@val} & \multicolumn{1}{c}{GAP} &\multicolumn{1}{c|}{Best AUC}& \multicolumn{1}{c}{PPL@val} & \multicolumn{1}{c}{GAP}&\multicolumn{1}{c}{Best AUC} \\ \midrule\midrule
LoRA & 1.098 & 0.059 & 0.735$\rightarrow$\textbf{0.765} & 1.041 & 0.006 & 0.687$\rightarrow$\textbf{0.721} & 0.950 & 0.021 & 0.689$\rightarrow$\textbf{0.775} \\
DoRA & 1.099 & 0.061 & 0.743$\rightarrow$\textbf{0.769} & 1.043 & 0.009 & 0.689$\rightarrow$\textbf{0.725} & 0.958 & 0.029 & 0.698$\rightarrow$\textbf{0.783} \\
qLoRA (Int8) & 1.100 & 0.057 & 0.722$\rightarrow$\textbf{0.751} & 1.044 & 0.007 & 0.683$\rightarrow$\textbf{0.722} & 0.951 & 0.021 & 0.686$\rightarrow$\textbf{0.773} \\
qLoRA (FP4) & 1.104 & 0.055 & 0.700$\rightarrow$\textbf{0.735} & 1.054 & 0.008 & 0.662$\rightarrow$\textbf{0.695} & 0.955 & 0.022 & 0.670$\rightarrow$\textbf{0.751}\\
\bottomrule
\end{tabular}}
\end{table*}

\section*{Impact of the Size of Pre-trained Model}

\begin{figure}[t]
\centering
\includegraphics[width=0.9\linewidth]{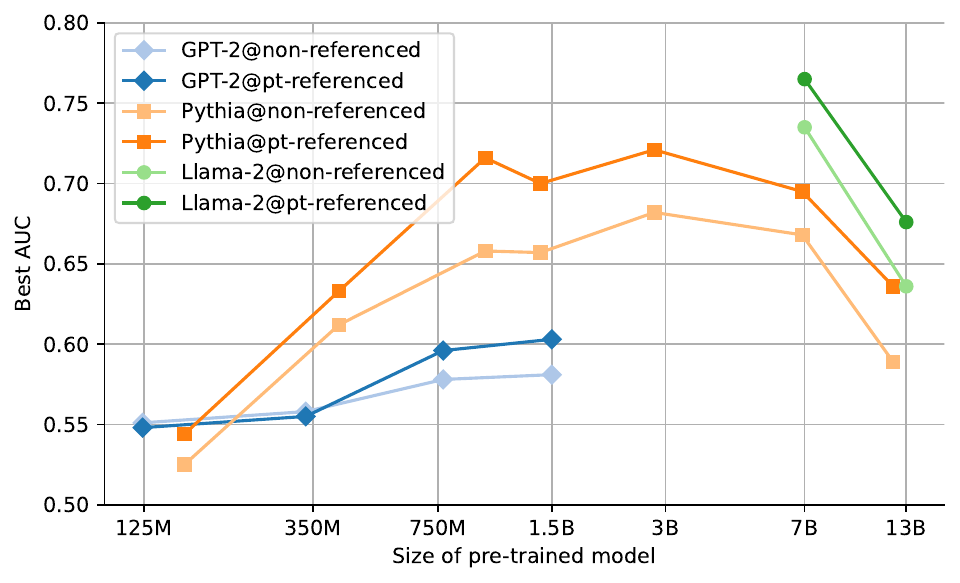}
\caption{The best AUC scores for non-referenced and pt-referenced MIAs on models fine-tuned on AG News, across various pre-trained model sizes.}
\label{fig:pre-train-size}
\end{figure}

To investigate the MIA risks associated with varying scales of pre-trained language models, we further fine-tune different sizes of pre-trained models on the AG News dataset using LoRA, including GPT-2 (124M, 335M, 774M, and 1.5B), Pythia (160M, 410M, 1B, 1.4B, 2.8B, 6.9B, and 12B), and Llama-2 (7B and 13B).
We follow the same fine-tuning settings as in~\Cref{sec:default-train} while selecting different rank $r$ so as to maintain approximately the same number of tuned parameters.
As illustrated in \Cref{fig:pre-train-size}, the MIA risk increases as the size of the pre-trained model scales up to around one billion (1B) parameters but decreases for models larger than this range.
This is because as the number of model parameters increases, the model's expressive power strengthens, leading to an increased degree of overfitting and thus ``memorizing'' more details of the fine-tuning data. 
This ``memorization'' heightens the risk of the model being susceptible to MIAs. 
However, once the model parameters reach a certain scale, the risk of overfitting may actually diminish, particularly when the model becomes more regularized or exhibits improved generalization capabilities.
Nevertheless, under all circumstances, pt-referenced MIAs expose the MIA risk more thoroughly than non-referenced MIAs in terms of best AUC scores.

\section*{Other PEFT Methods}
\label{sec:peft}
While our work primarily focuses on LoRA-based methods, we discuss the applicability of \leak to other PEFT methods in this section.
We fine-tune the Llama-2 model on AG News using both Prompt Tuning~\cite{lester-etal-2021-power} and IA3~\cite{NEURIPS2022_0cde695b}.
For Prompt Tuning, we set the number of virtual prompts as 20 tokens, initialized with ``Predict the topic of this news is World, Sports, Business or Sci/Tech''.
The best AUCs for non-referenced and pt-reference attacks are 0.511 and 0.546, respectively.
For IA3, we add the learned vectors to all attention and feed-forward layers.
The best AUC for non-referenced and pt-reference attacks are 0.517 and 0.551, respectively.
Overall, the risks associated with these PEFT methods remain relatively minor.
This may be due to the very small parameter sizes tuned. 
For instance, Prompt Tuning and IA3 only tuned 82k and 1.5M parameters, respectively, which is far less than LoRA with 10M parameters.
Therefore, although \leak does expose more MIA risks in these PEFT methods, their vulnerability to MIA remains minimal.

\begin{table*}[t]
\caption{
Results on dropout defense.
We report the best MIA AUC results (non-referenced$\rightarrow$pt-referenced) with the varying dropout rates ($\eta$).
}
\label{tab:defense-dropout}
\centering

\begin{tabular}{c|crc|crc|crc}
\toprule
 \multirow{2}{*}{$\eta$} & \multicolumn{3}{c|}{AG News} & \multicolumn{3}{c|}{OAsst} & \multicolumn{3}{c}{MedQA} \\\cmidrule(lr){2-4}  \cmidrule(lr){5-7} \cmidrule(lr){8-10}
~ & \multicolumn{1}{c}{PPL@val} & \multicolumn{1}{c}{GAP} &\multicolumn{1}{c|}{Best AUC}& \multicolumn{1}{c}{PPL@val} & \multicolumn{1}{c}{GAP} &\multicolumn{1}{c|}{Best AUC}& \multicolumn{1}{c}{PPL@val} & \multicolumn{1}{c}{GAP}&\multicolumn{1}{c}{Best AUC} \\ \midrule\midrule
w/o & 1.099 & 0.059 & 0.735$\rightarrow$\textbf{0.764} & 1.043 & 0.009 & 0.689$\rightarrow$\textbf{0.724} & 0.952 & 0.024 & 0.692$\rightarrow$\textbf{0.779} \\\midrule
0.05 & 1.098 & 0.059 & 0.736$\rightarrow$\textbf{0.765} & 1.041 & 0.007 & 0.687$\rightarrow$\textbf{0.721} & 0.950 & 0.021 & 0.689$\rightarrow$\textbf{0.775} \\
0.25 & 1.098 & 0.056 & 0.730$\rightarrow$\textbf{0.758} & 1.043 & 0.007 & 0.681$\rightarrow$\textbf{0.715} & 0.953 & 0.023 & 0.689$\rightarrow$\textbf{0.776} \\
0.45 & 1.097 & 0.051 & 0.714$\rightarrow$\textbf{0.741} & 1.043 & 0.005 & 0.673$\rightarrow$\textbf{0.711} & 0.953 & 0.022 & 0.682$\rightarrow$\textbf{0.769} \\
0.65 & 1.097 & 0.046 & 0.697$\rightarrow$\textbf{0.724} & 1.042 & 0.001 & 0.656$\rightarrow$\textbf{0.693} & 0.954 & 0.020 & 0.676$\rightarrow$\textbf{0.759} \\
0.85 & 1.096 & 0.033 & 0.661$\rightarrow$\textbf{0.686} & 1.042 & -0.007 & 0.626$\rightarrow$\textbf{0.663} & 0.955 & 0.016 & 0.655$\rightarrow$\textbf{0.736} \\
0.95 & 1.098 & 0.014 & 0.595$\rightarrow$\textbf{0.610} & 1.043 & -0.018 & 0.580$\rightarrow$\textbf{0.611} & 0.959 & 0.009 & 0.621$\rightarrow$\textbf{0.691} \\
0.99 & 1.108 & -0.003 & 0.543$\rightarrow$\textbf{0.553} & 1.050 & -0.030 & 0.523$\rightarrow$\textbf{0.543} & 0.970 & -0.006 & 0.555$\rightarrow$\textbf{0.602}\\
\bottomrule
\end{tabular}
\end{table*}

\begin{table*}[t]
\centering
\caption{The AUC of different MIAs against models fine-tuned from GPT-2 and Pythia. 
We highlight the results of the \colorbox{blue!10}{pt-referenced attacks}.}
\label{tab:main}
\begin{tabular}{c|ccc|ccc}
\toprule
\multirow{2}{*}{Attack} & \multicolumn{3}{c}{GPT-2} & \multicolumn{3}{|c}{Pythia}   \\ \cmidrule(l){2-4}\cmidrule(l){5-7}
 & AG News & OAsst & MedQA & AG News & OAsst & MedQA \\ 
 \midrule \midrule
zlib & 0.549 & 0.514 & 0.536 & 0.609 & 0.520 & 0.536 \\
\midrule
GradNorm$_\theta$ & 0.568 & 0.512 & 0.545 & 0.626 & 0.533 & 0.590 \\
\midrule
LOSS & 0.558 & 0.501 & 0.544 & 0.620 & 0.508 & 0.579 \\
\cellcolor{blue!10}{~~~~~~~~~~$\hookrightarrow$+\textit{Pre}} & \cellcolor{blue!10}{0.578} & \cellcolor{blue!10}{0.524} & \cellcolor{blue!10}{0.547} & \cellcolor{blue!10}{0.661} & \cellcolor{blue!10}{0.555} & \cellcolor{blue!10}{0.583} \\
\midrule
Neighborhood & 0.543 & 0.496 & 0.519 & 0.599 & 0.497 & 0.536 \\
\cellcolor{blue!10}{~~~~~~~~~~$\hookrightarrow$+\textit{Pre}} & \cellcolor{blue!10}{0.566} & \cellcolor{blue!10}{0.523} & \cellcolor{blue!10}{0.542} & \cellcolor{blue!10}{0.653} & \cellcolor{blue!10}{0.592} & \cellcolor{blue!10}{0.585} \\
\midrule
Min-K\% & 0.559 & 0.510 & 0.548 & 0.629 & 0.525 & 0.592 \\
\cellcolor{blue!10}{~~~~~~~~~~$\hookrightarrow$+\textit{Pre}} & \cellcolor{blue!10}{0.586} & \cellcolor{blue!10}{0.531} & \cellcolor{blue!10}{0.558} & \cellcolor{blue!10}{0.678} & \cellcolor{blue!10}{0.564} & \cellcolor{blue!10}{0.611} \\
\midrule
Min-K\%++ & 0.570 & 0.528 & 0.569 & 0.682 & 0.599 & 0.633 \\
\cellcolor{blue!10}{~~~~~~~~~~$\hookrightarrow$+\textit{Pre}} & \cellcolor{blue!10}{0.603} & \cellcolor{blue!10}{0.517} & \cellcolor{blue!10}{0.566} & \cellcolor{blue!10}{0.721} & \cellcolor{blue!10}{0.606} & \cellcolor{blue!10}{0.686} \\
\midrule
MoPe & 0.536 & 0.519 & 0.550 & 0.586 & 0.514 & 0.558 \\
\cellcolor{blue!10}{~~~~~~~~~~$\hookrightarrow$+\textit{Pre}} & \cellcolor{blue!10}{0.552} & \cellcolor{blue!10}{0.546} & \cellcolor{blue!10}{0.592} & \cellcolor{blue!10}{0.674} & \cellcolor{blue!10}{0.643} & \cellcolor{blue!10}{0.650}  \\
\midrule
GradNorm$_x$ & 0.581 & 0.540 & 0.555 & 0.630 & 0.547 & 0.602 \\
\cellcolor{blue!10}{~~~~~~~~~~$\hookrightarrow$+\textit{Pre}} & \cellcolor{blue!10}{0.566} & \cellcolor{blue!10}{0.539} & \cellcolor{blue!10}{0.541} & \cellcolor{blue!10}{0.606} & \cellcolor{blue!10}{0.576} & \cellcolor{blue!10}{0.599} \\
\bottomrule
\end{tabular}
\end{table*}

\begin{figure*}[t]
    \centering
    \includegraphics[width=1\linewidth]{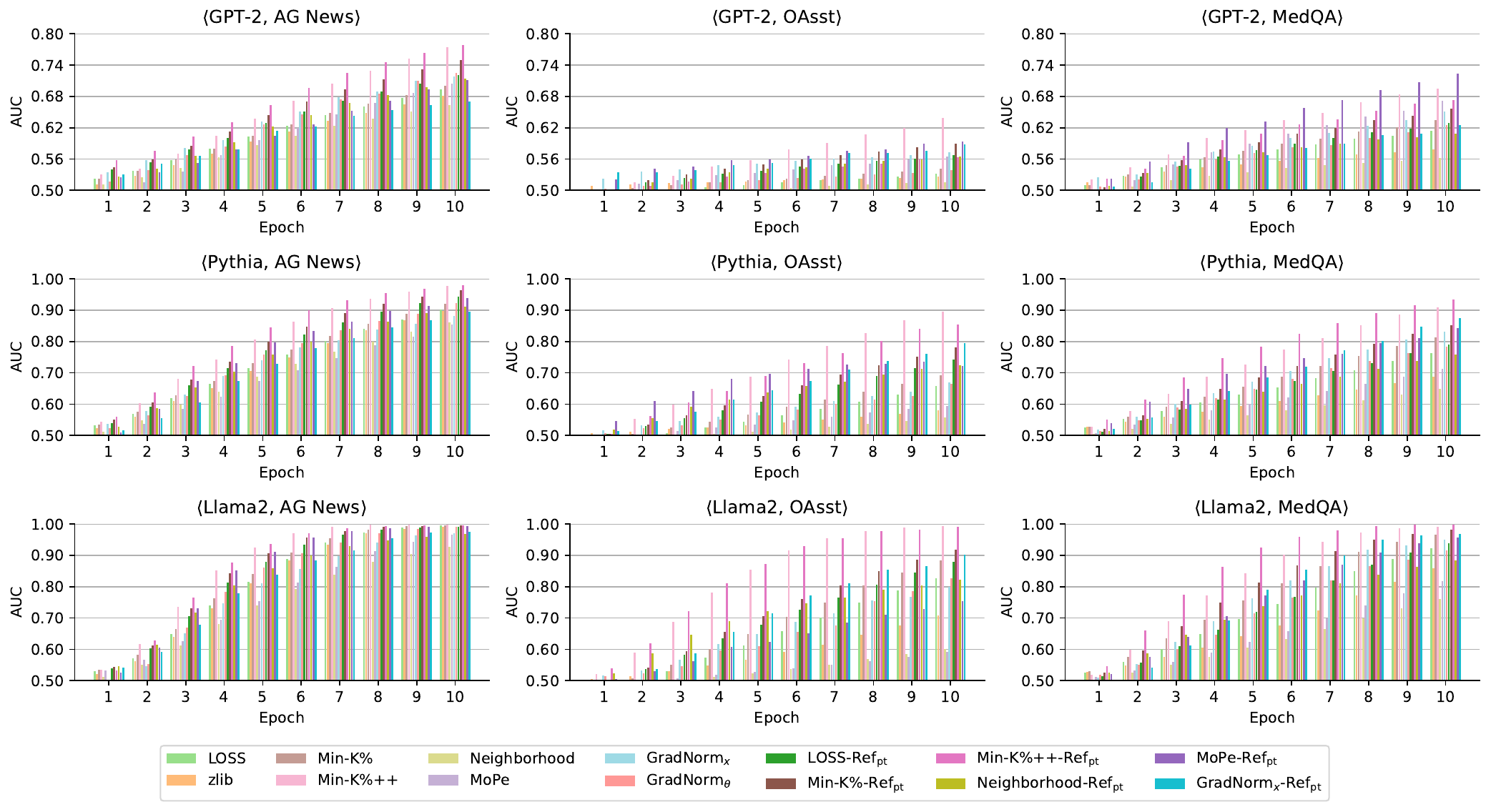}
    \caption{The AUC of various MIAs when fine-tuning the pre-trained models with different \underline{epochs}.}
    \label{fig:each-epochs}
\end{figure*}

\begin{figure*}[t]
    \centering
    \includegraphics[width=1\linewidth]{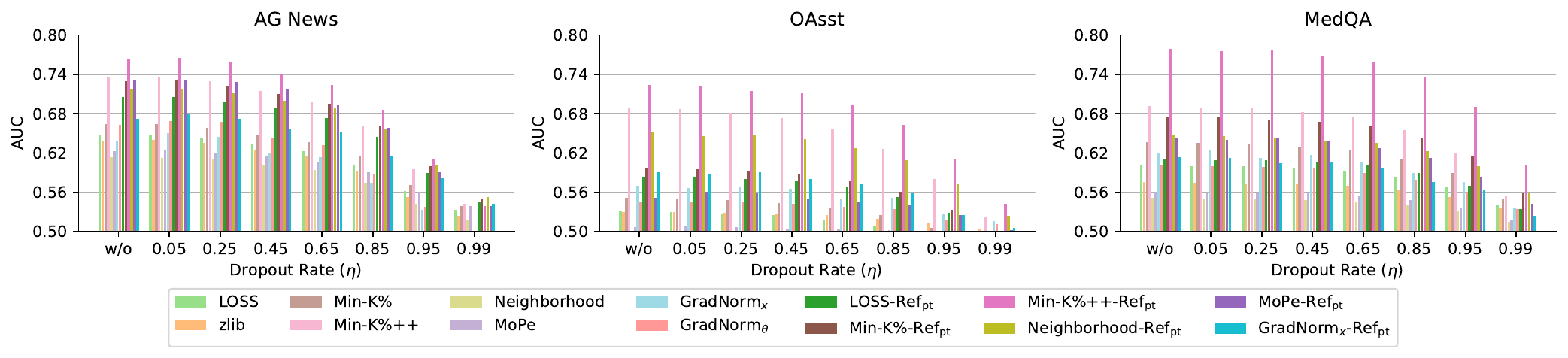}
    \caption{The AUC of various MIAs against the fine-tuned Llama-2 with varying \underline{dropout rates ($\eta$)}.}
    \label{fig:each-dropout}
\end{figure*}

\begin{figure*}[t]
    \centering
    \includegraphics[width=1\linewidth]{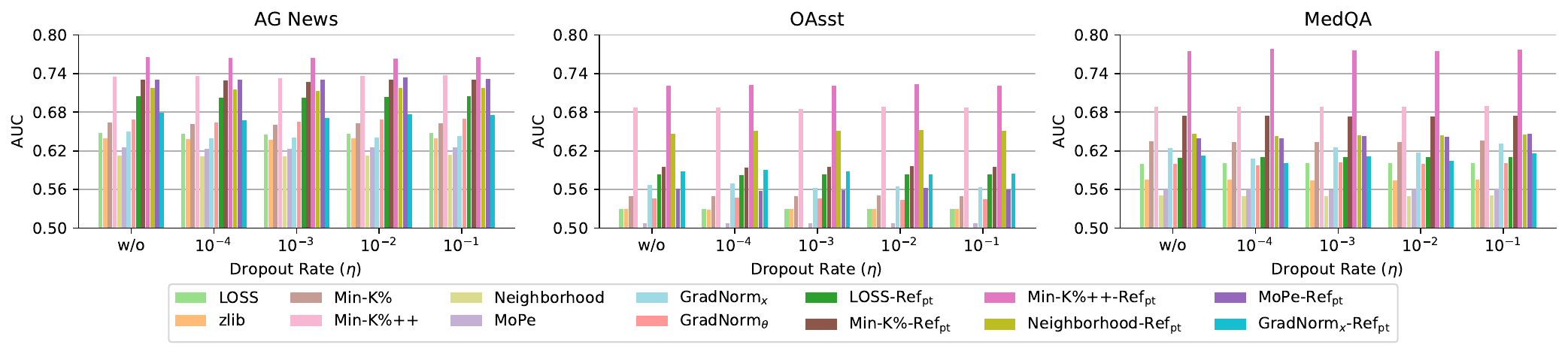}
    \caption{The AUC of various MIAs against the fine-tuned Llama-2 with varying \underline{weight decay rates ($\alpha$)}.}
    \label{fig:each-weight}
\end{figure*}

\begin{figure*}[t]
\centering
\includegraphics[width=1\linewidth]{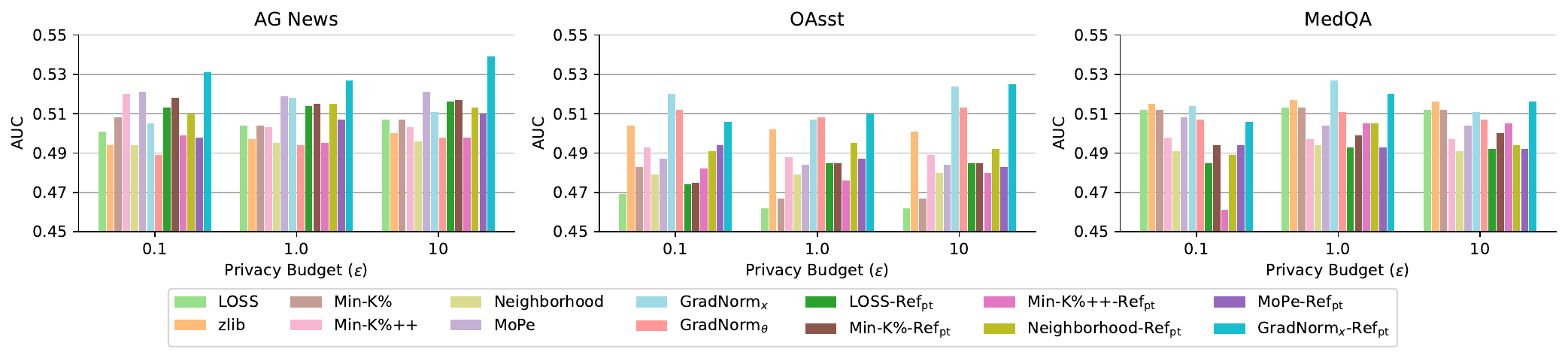}
\caption{The AUC of various MIAs against the fine-tuned Llama-2 with varying \underline{privacy budgets ($\epsilon$)}.}
\label{fig:each-priv-budget}
\end{figure*}

\begin{figure*}[t]
\centering
\includegraphics[width=1\linewidth]{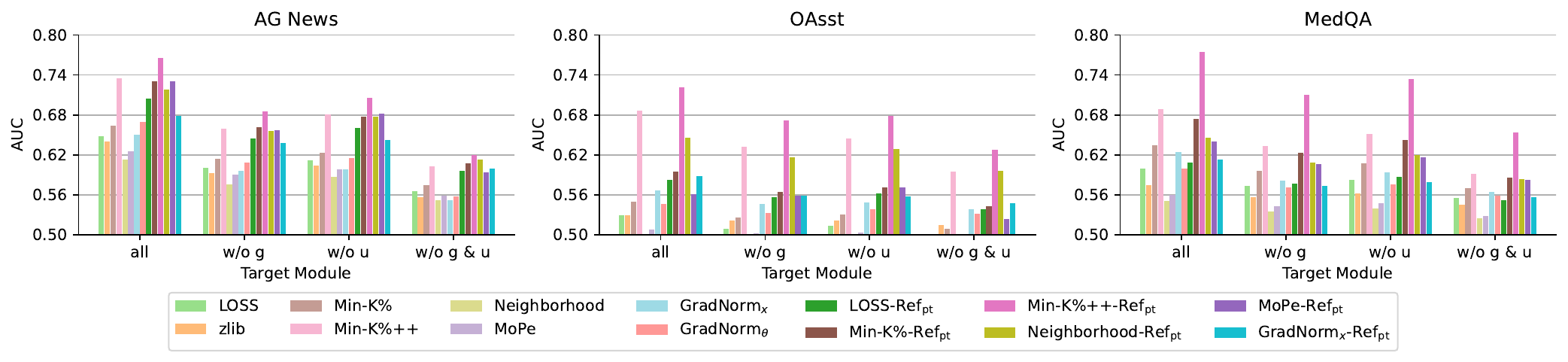}
\caption{The AUC of various MIAs against the fine-tuned Llama-2 \underline{with and without the \textit{up} and \textit{gate} layers}.}
\label{fig:each-specific-layer}
\end{figure*}

\begin{figure*}[t]
\centering
\includegraphics[width=0.9\linewidth]{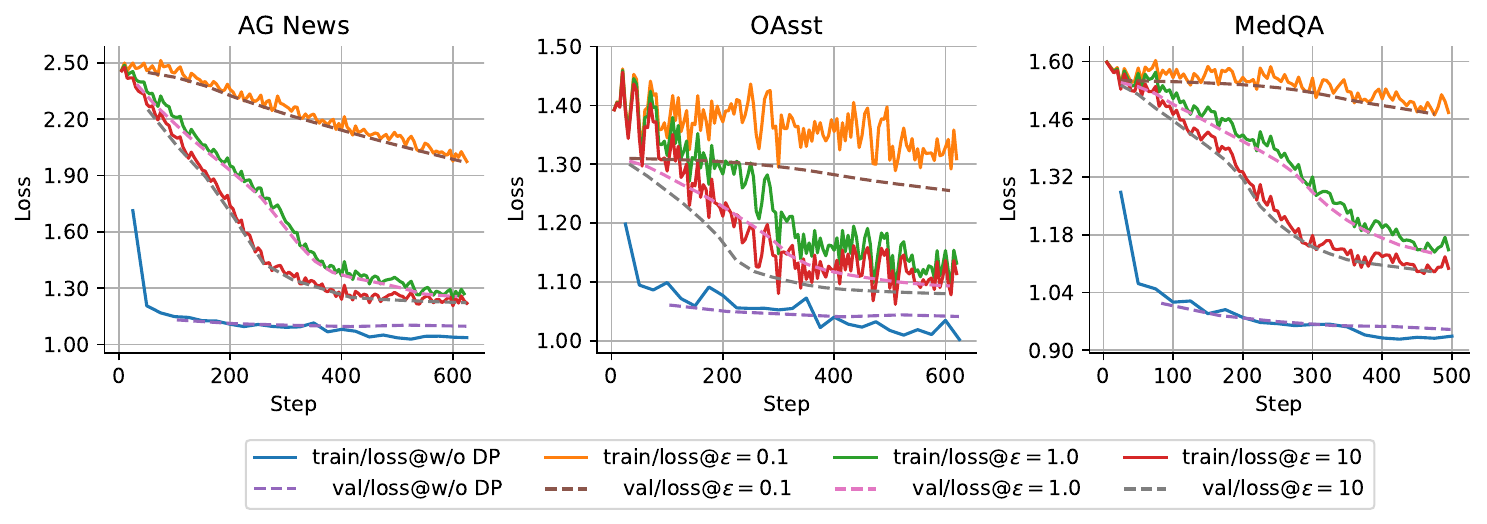}
\caption{The training and validation loss during Llama-2 LoRA fine-tuning across three datasets when applying differential privacy with varying privacy budgets $\epsilon$.}
\label{fig:ft-dp}
\end{figure*}

\end{document}